\documentclass[preprint]{aastex}

\newcommand{\kms}{km~s$^{-1}$}
\newcommand{\myr}{M$_\odot$~yr$^{-1}$}
\newcommand{\whz}{W~Hz$^{-1}$}

\slugcomment{To appear in the {\it Astronomical Journal}}

\begin{document}

\title{The Radio Luminosity Function and Galaxy Evolution in the Coma Cluster}

\author{Neal A. Miller\altaffilmark{1}} 
\email{nmiller@pha.jhu.edu}

\author{Ann E. Hornschemeier\altaffilmark{1,2}}

\author{Bahram Mobasher\altaffilmark{3}}

\author{Terry J. Bridges\altaffilmark{4,5}}

\author{Michael J. Hudson\altaffilmark{6}}

\author{Ronald O. Marzke\altaffilmark{7}}

\author{Russell J. Smith\altaffilmark{8}}

\altaffiltext{1}{Department of Physics and Astronomy, Johns Hopkins University, 3400 N. Charles Street, Baltimore, MD 21218}
\altaffiltext{2}{Laboratory for X-Ray Astrophysics, NASA Goddard Space Flight Center, Code 662.0, Greenbelt, MD 20771}
\altaffiltext{3}{Department of Physics and Astronomy, University of California, Riverside, CA 92521}
\altaffiltext{4}{Anglo-Australian Observatory, P.O. Box 296, Epping, NSW 1710, Australia}
\altaffiltext{5}{Department of Physics, Queen's University, Kingston, Ontario K7L 3N6, Canada}
\altaffiltext{6}{Department of Physics and Astronomy, University of Waterloo, 200 University Avenue West, Waterloo, Ontario, N2L 3G1, Canada}
\altaffiltext{7}{Department of Physics and Astronomy, San Francisco State University, San Francisco, CA 94132}
\altaffiltext{8}{Department of Physics, University of Durham, Durham DH1 3LE, UK}

\begin{abstract} 
We investigate the radio luminosity function and radio source population for two fields within the Coma cluster of galaxies, with the fields centered on the cluster core and southwest infall region and each covering about half a square degree. Using VLA data with a typical rms sensitivity of 28 $\mu$Jy per 4\farcs4 beam, we identify 249 radio sources with optical counterparts brighter than $r = 22$. For cluster galaxies, these correspond to $L_{1.4} = 1.7 \times 10^{20}$ \whz{}(for a 5$\sigma$ source) and $M_r = -13$. Comprehensive optical spectroscopy identifies 38 of these as members of the Coma cluster, evenly split between sources powered by an active nucleus and sources powered by active star formation. The radio-detected star-forming galaxies are the dominant population only at radio luminosities between about $10^{21}$ and $10^{22}$ \whz, an interesting result given that star formation dominates field radio luminosity functions for {\em all} luminosities lower than about $10^{23}$ \whz. The majority of the radio-detected star-forming galaxies have characteristics of starbursts, including high specific star formation rates and optical spectra with strong emission lines. In conjunction with prior studies on post-starburst galaxies within the Coma cluster, this is consistent with a picture in which late-type galaxies entering Coma undergo a starburst prior to a rapid cessation of star formation. Optically bright elliptical galaxies ($M_r \leq -20.5$) make the largest contribution to the radio luminosity function at both the high ($\gtrsim 3 \times 10^{22}$ \whz) and low ($\lesssim 10^{21}$ \whz) ends. Through a stacking analysis of these optically-bright ellipticals we find that they continue to harbor radio sources down to luminosities as faint as $3 \times 10^{19}$ \whz. However, contrary to published results for the Virgo cluster we find no evidence for the existence of a population of optically faint ($M_r \approx -14$) dwarf ellipticals hosting strong radio AGN.

\end{abstract}

\section{Introduction}\label{sec-intro}

The radio population of galaxies is composed of a mix of sources powered by active galactic nuclei (AGN) and sources whose radio emission traces its origin to star formation. Locally-determined radio luminosity functions \citep[RLFs; e.g.,][]{condon2002,mauch2007} demonstrate that nearly all sources with $L_{1.4GHz} \geq 10^{23}$ \whz{} are AGN, while below this 1.4~GHz luminosity AGN continue to be a significant population although sources powered by star formation become progressively more numerous than those powered by AGN. The AGN at brighter radio luminosities are the sources typically thought of as ``radio galaxies,'' with radio emission extended in jets or lobes often beyond the optical extent of the host galaxy. For star-forming galaxies, the radio emission arises from the synchrotron radiation of electrons accelerated by the supernovae of massive stars, and thus is a near-current indicator of the star formation rate \citep[e.g.,][]{condon1992}. 

The construction of these previous RLFs was made possible by the availability of wide-field radio surveys such as the NRAO\footnote{The National Radio Astronomy Observatory is a facility of the National Science Foundation operated under cooperative agreement by Associated Universities, Inc.} VLA Sky Survey \citep[NVSS;][]{condon1998} in conjunction with local galaxy catalogs and redshift surveys. The large areal coverage means that even with the modest flux density limit of the NVSS (2.5 mJy) RLFs can be constructed down to faint 1.4~GHz luminosities, around $10^{20}$ \whz. For the case of star-forming galaxies, this translates to low star formation rates (SFR) of about 0.06 \myr \citep[using the conversion of ][which assumes a Salpeter IMF and includes all stars with 0.1M$_\odot$ $\leq$ M $\leq$ 100M$_\odot$]{yun2001}.

Local environment does affect the RLF, as evidenced in galaxy clusters. Based on a sample of 20 nearby galaxy clusters, \citet{miller2002} found that the crossover point where star formation dominates the RLF occurs at lower radio luminosities ($L_{1.4GHZ} \approx 5 \times 10^{22}$ \whz). Presumably this is the result of the different morphological mix of galaxies in clusters compared to the field; the morphology-density relation \citep[e.g.,][]{dressler1980,postman1984,goto2003,thomas2006} implies a larger fraction of massive ellipticals in clusters and these sources are rarely associated with star formation. Running parallel to this argument are the numerous environmental effects that suppress star formation in clusters and the observed density-SFR relation \citep{gomez2003}, further reducing the relative contribution of star-forming galaxies to the cluster RLF. 

However, cluster RLF work has generally explored only the brighter radio luminosities. The \citet{miller2002} cluster RLF is complete only to around $5 \times 10^{21}$ \whz, a value roughly consistent with the 1.4~GHz luminosity of the Milky Way galaxy and equivalent to a SFR of around 3 \myr. Smaller galaxies are more prone to environmental effects, and hence exploration of lower 1.4~GHz luminosities can reveal more dramatic variation in radio galaxy populations. \citet[][hereafter GB99]{gavazzi1999} correlated the NVSS with the Virgo Cluster Catalog of \citet[][]{binggeli1985} to explore the RLF and faint radio source populations. Surprisingly they identified very optically faint dwarf ellipticals ($M_B \leq -13$) hosting relatively powerful ($L_{1.4GHz} \approx 10^{21}$ \whz) radio sources. In fact, the probability for a $M_B = -13$ dwarf elliptical to host a strong radio source was found to be equivalent to that of a dwarf elliptical three to four magnitudes brighter. This is in contradiction to findings for massive elliptical galaxies where the probability for an object to be associated with a radio source is a strong function of optical luminosity \citep{auriemma1977,ledlow1996}. GB99 assume that these fainter dwarf ellipticals are AGN, and it is relatively simple to exclude star formation as the cause of their radio emission. Converting from magnitude to luminosity and applying a mass-to-luminosity ratio of 2 \citep[e.g., equation 10 of ][]{magorrian1998}, an $M_B = -13$ elliptical would have a mass of under $10^{8}$ M$_\odot$. The SFR implied by a radio luminosity of $10^{21}$ \whz{} is 0.6 \myr, meaning a very high specific star formation rate capable of assembling the entire mass of the galaxy in a starburst episode of less than 200 million years. The colors of these dwarf galaxies are not consistent with recent star formation and presumably such objects would also be obvious in the ultraviolet and optical emission line surveys.

In this paper, we investigate the RLF and radio source population of the Coma cluster. The radio data correspond to a new VLA survey that is able to detect Coma galaxies with 1.4~GHz luminosities as low as $1.3 \times 10^{20}$ \whz{} \citep[][hereafter MHM2008]{miller2008}. The relative proximity of Coma \citep[$z=0.0231$, e.g.][]{struble1999} means that the Sloan Digital Sky Survey \citep[SDSS;][]{york2000} provides optical identifications down to $M_r \approx -13$, along with optical spectroscopy of galaxies as faint as $M_r \approx -17$. Along with other published redshift surveys and a new MMT Hectospec survey of Coma (Marzke et al., in preparation), the spectroscopic completeness for the radio-optical identifications is outstanding. We are thus able to assess the RLF and source population of the Coma cluster, including possible sources with very faint optical magnitudes. 

Specifically, we seek to address the composition of the faint radio galaxy population of Coma. What are the relative contributions of star-forming galaxies and AGN to the faint end of the Coma RLF? How do these compare to RLFs constructed for field galaxies, and what might be the cause of any such differences? Finally, is there evidence for powerful AGN in optically-faint dwarf elliptical galaxies? 

We describe the data in Section \ref{sec-data}, including an overview of the radio observations and their matching to optical photometry and spectroscopy. We then detail the properties of the radio-detected members in Section \ref{sec-pops} and divide them into those with radio emission primarily associated with an active nucleus and those associated with star formation. This information is used to construct the RLF and its contributions from each type of source (Section \ref{sec-rlf}). We discuss the results in Section \ref{sec-discuss}, including the composition of the RLF at low luminosities (Section \ref{sec-declinesf}), the prevalence of starbursts among star-forming galaxies (Section \ref{sec-bursts}), and AGN in cluster ellipticals from the brighter such galaxies on down to the dwarfs explored by GB99 (Section \ref{sec-ellips}). For all calculations that require it we have assumed that the distance to Coma is 100 Mpc, which corresponds to a distance modulus of $m-M = 35$ magnitudes.

\section{The Radio Galaxy Sample}\label{sec-data}

\subsection{Radio Data and Optical IDs}

The radio observations and data reduction are summarized in MHM2008. Briefly, a mosaic of VLA pointings was used to produce near-uniform sensitivity images of a pair of 32\farcm5 $\times$ 50\farcm8 fields: ``Coma 1'' with a coordinate center of (J2000) 12$^{\mbox{{\scriptsize h}}}$59$^{\mbox{{\scriptsize m}}}$45\fs2 $+$27$^\circ$57\arcmin53\farcs1, and ``Coma 3'' centered on 12$^{\mbox{{\scriptsize h}}}$57$^{\mbox{{\scriptsize m}}}$28\fs5 $+$27$^\circ$08\arcmin08\farcs5 \citep{komiyama2002}. The southwest corner of Coma 1 overlaps slightly with the northeast corner of Coma 3, such that when combined they define a contiguous total area of about 0.92 square degrees. The rms sensitivity across this full area is about 28 $\mu$Jy per 4\farcs4 beam, with the deepest regions having an rms noise of 22 $\mu$Jy per beam. This corresponds to a 5$\sigma$ detection of as low as $1.3 \times 10^{20}$ \whz{} for point sources in Coma. The 4\farcs4 resolution of the radio data corresponds to 2 kpc at the assumed distance of 100 Mpc.

Although the VLA data and catalog cover a much larger area, we restrict our analysis to the region described by the Coma 1 and Coma 3 fields. There are two main reasons for adopting this restricted area. First, the sensitivity within this area is fairly uniform and well defined, allowing us to apply small area corrections to counts made at lower flux densities. Second, the Coma 1 and Coma 3 regions have been the focus of other multiwavelength observational campaigns, most notably optical spectroscopy \citep[e.g.,][Marzke et al., in preparation]{mobasher2001}. We will subsequently often refer to the combined Coma 1 and Coma 3 fields as the survey area. Within the survey area, there are a total of 628 radio sources detected at 5$\sigma$ or greater.

Optical counterparts to the radio sources were identified from the SDSS sixth data release \citep[DR6;][]{adelman2008} as detailed in MHM2008. The SDSS 95\% completeness limit for point sources in typical seeing is $r = 22.2$ \citep{adelman2007}, with star-galaxy separation accurate to $r \sim 21.5$ \citep{scranton2002}. Consequently we consider only radio sources with optical counterparts of $r \leq 22$ in this study, and accept such sources when they lie within 2\arcsec{} of the radio position. A small number of sources with larger separations were accepted on the basis of their radio and optical morphologies. Within the survey area there are a total of 248 radio sources with non-stellar (SDSS PhotoType = 3, galaxy), $r \leq 22$ optical counterparts. This represents approximately 40\% of the full list of radio detections, and including SDSS sources with stellar classifications as well as those with $r > 22$ increases this fraction to over 50\%. We add one further source associated with a lower surface brightness radio detection (see Section \ref{sec-adds}). Of the 249 total optical-radio IDs, about five are expected to be chance superpositions as determined through numerous offsets of the radio catalog relative to the optical data (see Figure 4 of MHM2008). The majority of these are for sources with $r > 20$. The magnitude distribution of the radio-optical sources is shown in Figure \ref{fig-specsamp}. Our criteria of using only SDSS non-stellar objects removed 18 stellar objects (SDSS PhotoType = 6) from consideration as possible Coma radio galaxies, only six of which have $r \leq 20$ with the brightest having $r=17.74$. 

\subsection{Optical Spectroscopy of Radio Galaxies}\label{sec-optspec}

Coma has been the target for numerous spectroscopic surveys, providing redshifts for many of the radio-detected galaxies \citep[e.g.,][]{caldwell1993,biviano1995,castander2001,mobasher2001}. We have used the NASA/IPAC Extragalactic Database (NED) to search for publicly-available velocity measurements for all of the radio-detected galaxies. In addition, the SDSS included spectroscopy of the Coma cluster in its sixth data release \citep[SDSS DR6;][]{adelman2008}. The SDSS main galaxy sample provides spectra of galaxies with $r \lesssim 17.8$ \citep{strauss2002}, although the high density of galaxies in Coma limits the total number of available SDSS spectra. We have also undertaken a campaign aimed at providing optical spectroscopy for fainter galaxies in the Coma cluster (Marzke et al., in preparation). The 2007 phase of this program used the MMT and its multi-fiber spectrograph (Hectospec) to observe galaxies down to $r \sim 21.3$ with no optical color bias. Fiber assignments included higher priority for objects detected in multiwavelength programs such as the VLA survey.

\begin{figure}
\figurenum{1}
\epsscale{0.9}
\plotone{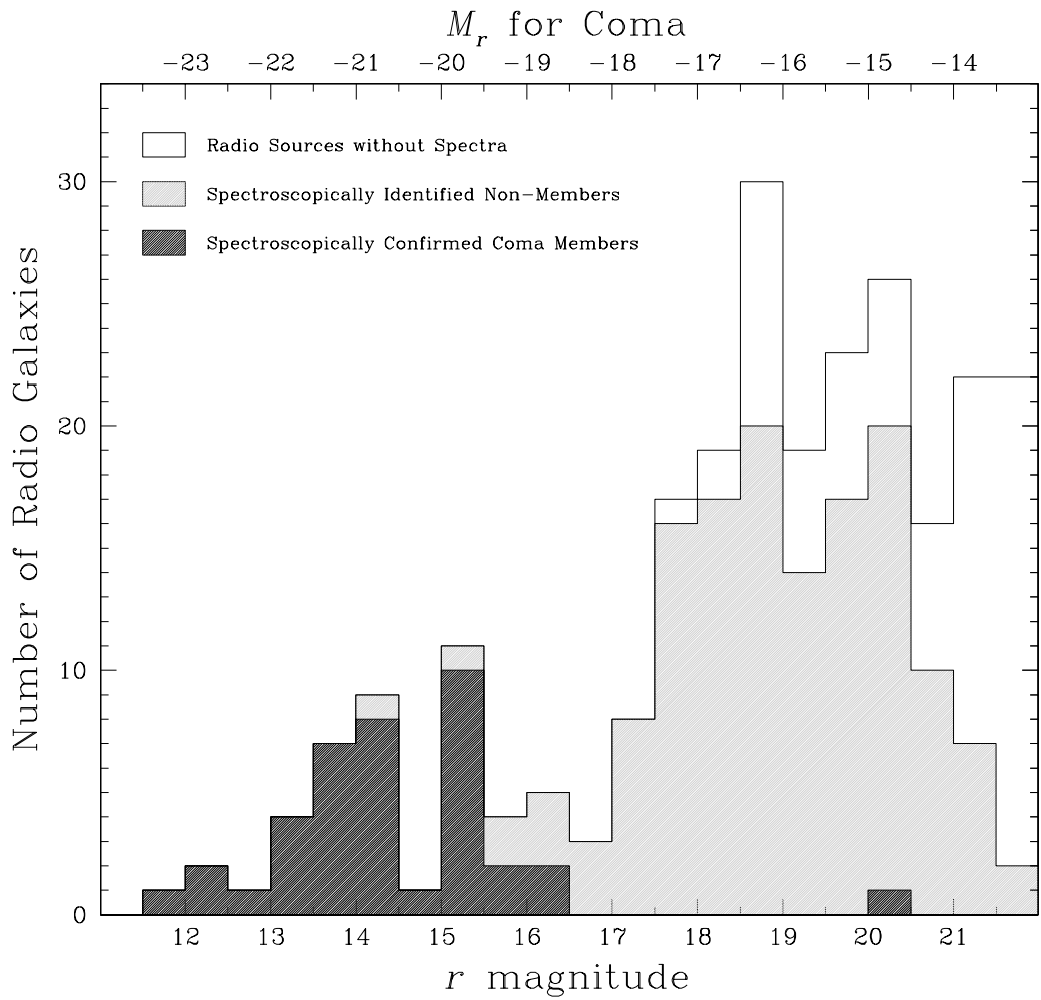}
% SM file is samp.sm, on A1656/Multi/Paper/ directory
\caption{Magnitude distribution and spectroscopic sampling for radio-detected galaxies.\label{fig-specsamp}}
\end{figure}

\begin{figure}
\figurenum{2}
\epsscale{0.9}
\plotone{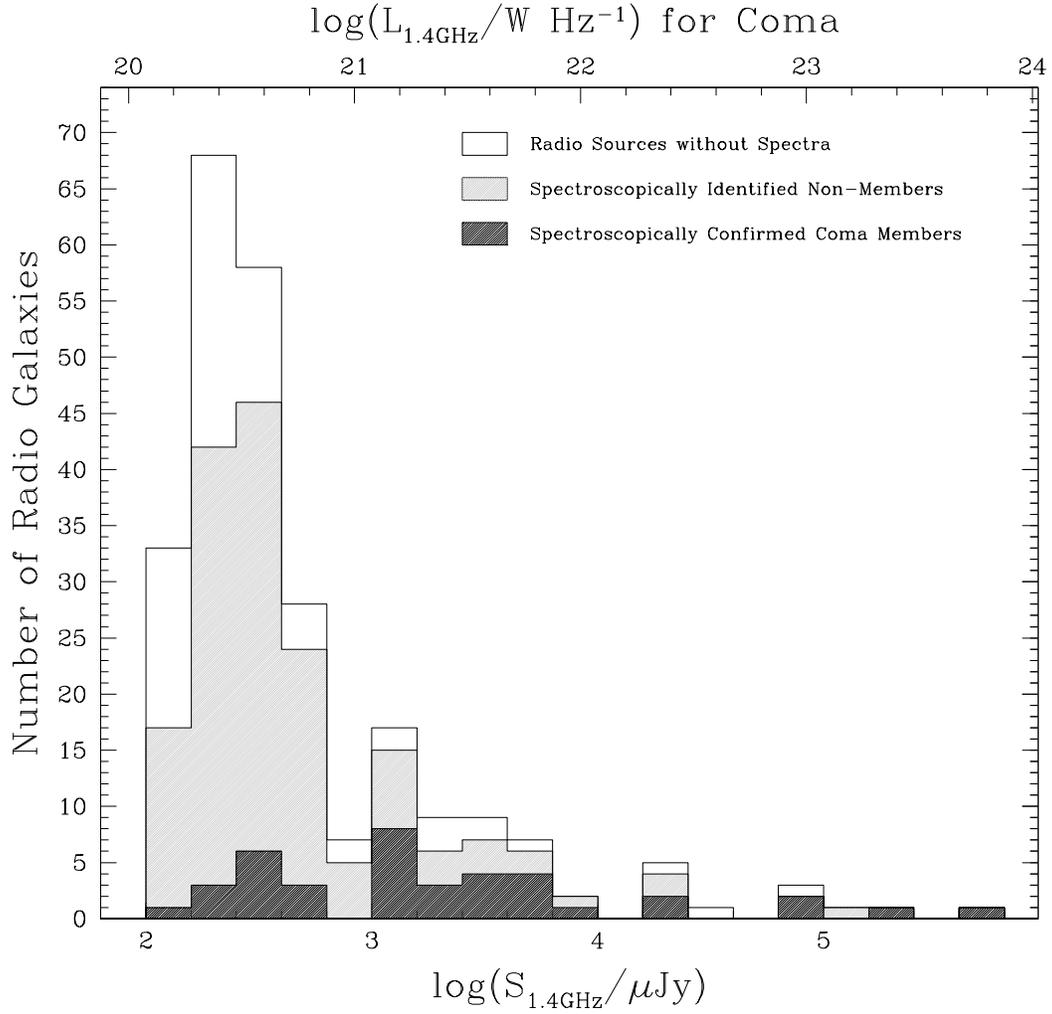}
\caption{Flux density distribution and spectroscopic sampling for radio-detected galaxies with $r \leq 22$.\label{fig-specsampradio}}
\end{figure}

\begin{figure}
\figurenum{3}
\epsscale{0.9}
\plotone{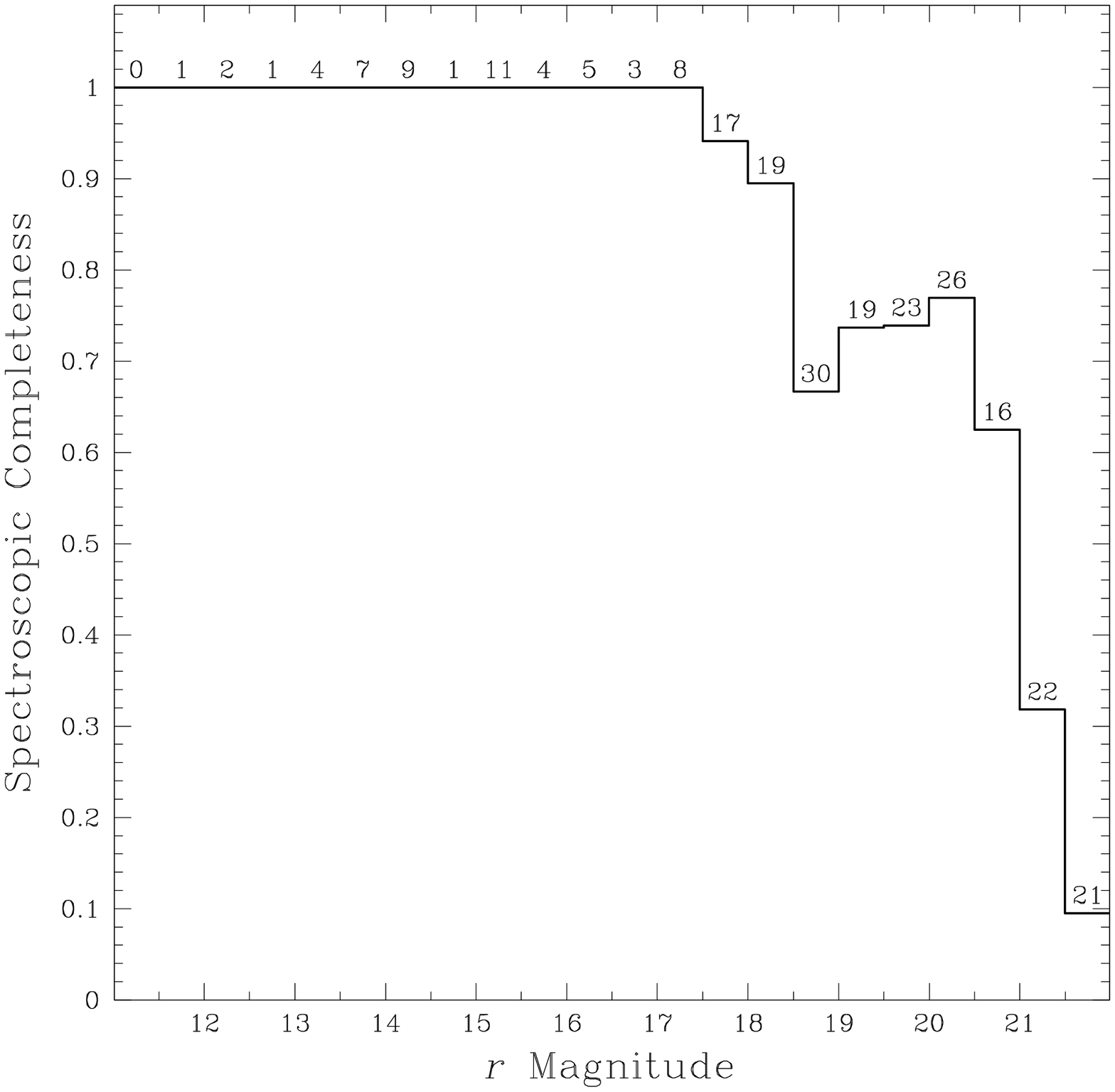}
\caption{Fractional completeness of spectroscopic sampling as a function of optical magnitude. The total number of objects within each bin is incidated, and bins which do not contain any objects have their completeness set to unity.\label{fig-spfraco}}
\end{figure}

\begin{figure}
\figurenum{4}
\epsscale{0.9}
\plotone{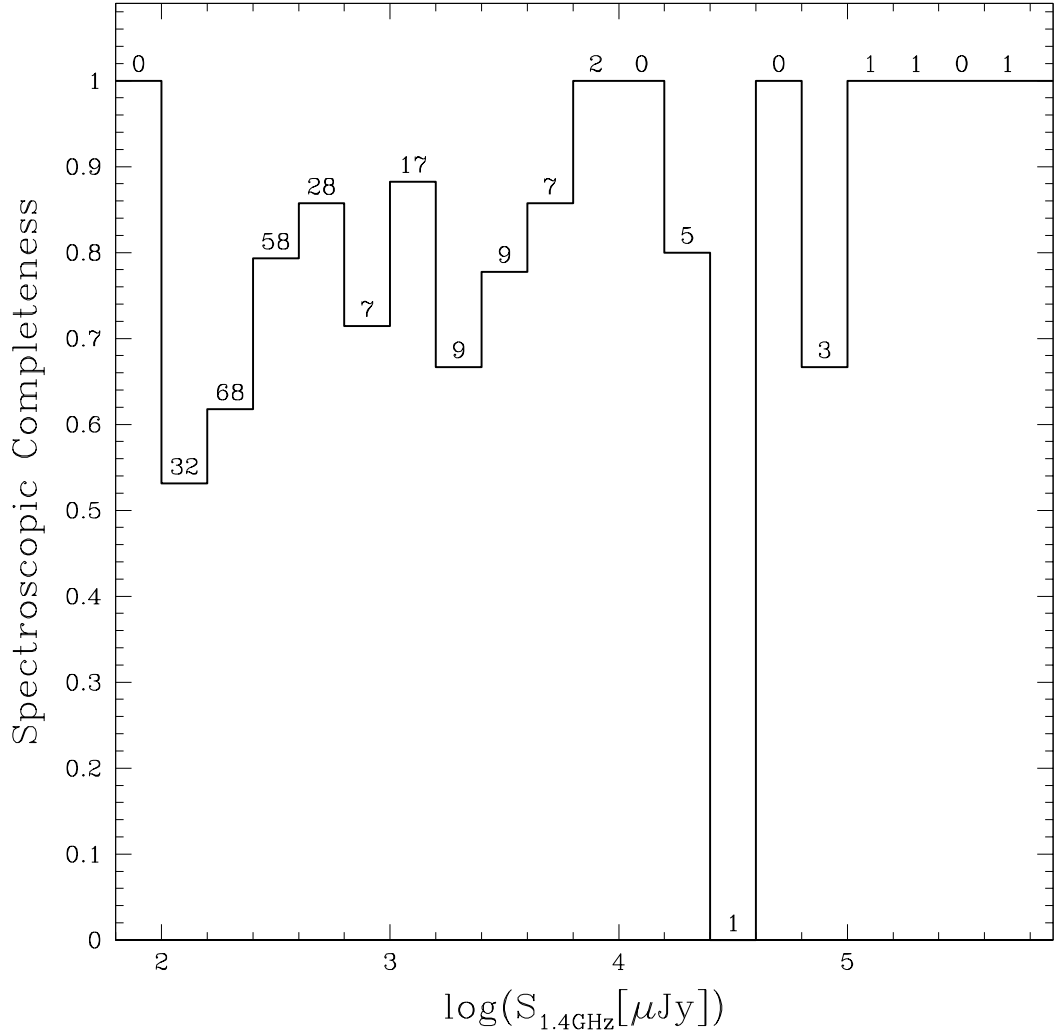}
\caption{Fractional completeness of spectroscopic sampling as a function of 1.4~GHz flux density. The total number of objects within each bin is indicated, and bins which do not contain any objects have their completeness set to unity.\label{fig-spfracr}}
\end{figure}

The spectroscopic completeness is shown in Figures \ref{fig-specsamp} - \ref{fig-spfracr}. In Figure \ref{fig-specsamp} we show a histogram of the optical magnitude distribution of the radio-detected galaxies, where shaded portions indicate sources with spectroscopic measurements. Figure \ref{fig-specsampradio} is the analogous histogram for the radio flux density distribution. The spectroscopic sampling implied by each of these figures is depicted in Figures \ref{fig-spfraco} and \ref{fig-spfracr}, respectively. In total, 179 of the 249 radio-detected galaxies have spectra with up to 39 of these belonging to the Coma cluster \citep[defined as 4000 \kms{}$\geq cz \leq$ 10000 \kms;][]{colless1996}. The brightest radio-detected galaxy without a spectroscopic measurement has $r=17.97$, meaning we are spectroscopically 100\% complete to about $M_r = -17$ (Figure \ref{fig-spfraco}). The spectroscopic completeness then hovers around 70\% to $r = 21$ ($M_r = -14.5$) before falling rapidly for fainter magnitudes. There is a similar trend for the spectroscopic completeness as a function of radio flux density with somewhere around 80\% of sources down to about 250 $\mu$Jy having attendant optical spectroscopy. Even at flux densities below this value more than half of the sources have spectroscopic measurements.

\begin{figure}
\figurenum{5}
\epsscale{0.9}
\plotone{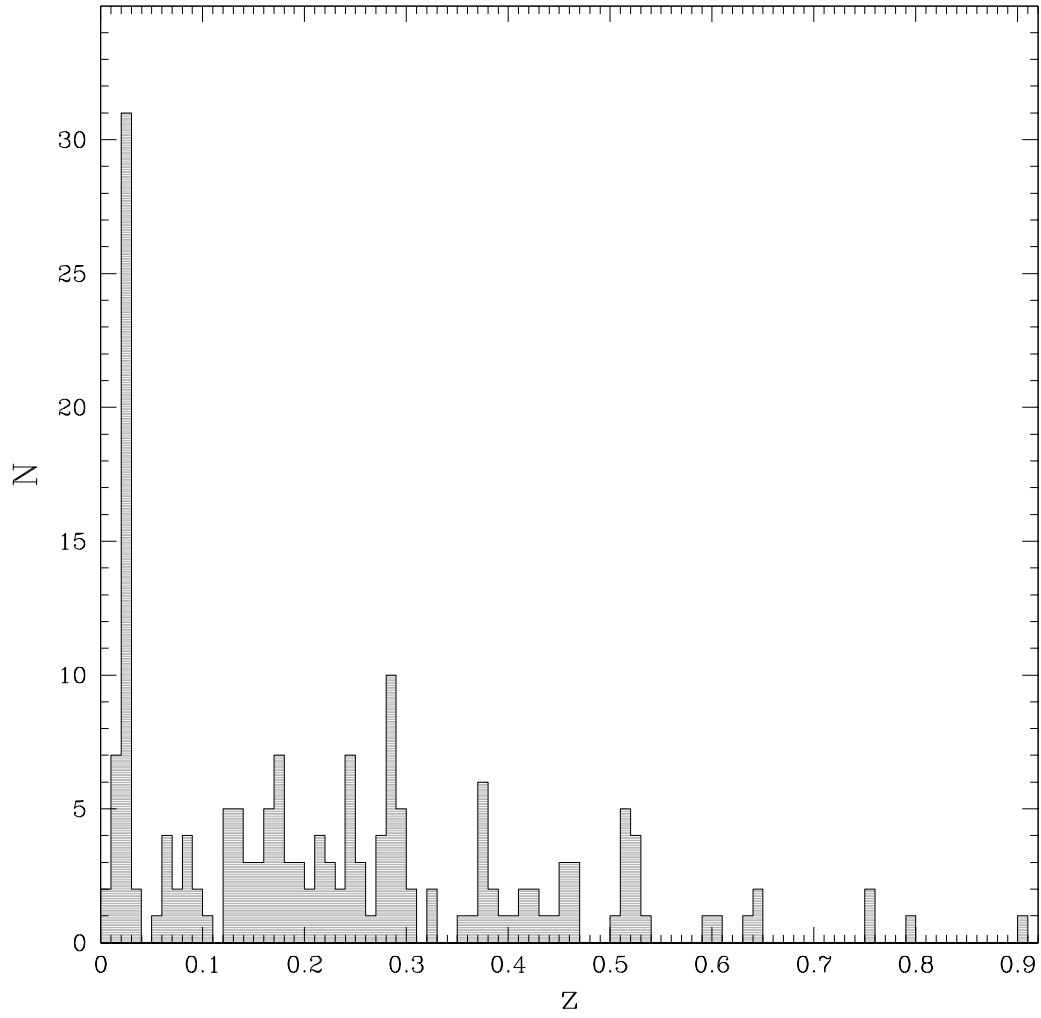}
\caption{Redshift histogram for the 179 radio-detected galaxies with optical spectra. The bin size is 0.01, or approximately 3,000 \kms.\label{fig-vhist}}
\end{figure}

Table \ref{tbl-mems} lists the Coma members and some of their properties, ordered by $r$ magnitude. The other 140 radio-detected objects with optical spectra almost exclusively lie behind Coma ($cz > 10000$ \kms), as shown in Figure \ref{fig-vhist}. One exception is the Galactic planetary nebula PG1257+279 \citep{green1986}, that has a photometric tag of galaxy in the SDSS since it is resolved. Velocities for the non-Coma members will be presented in Marzke et al. (in preparation). The $r$ magnitude distribution of radio-detected members of the Coma cluster (Figure \ref{fig-specsamp}) includes one obvious outlier: J12591536+2746052, which is nearly four magnitudes fainter than the next-faintest radio-detected galaxy in Coma. The source of its velocity measurement is \citet{biviano1995}, who indicated that this determination is only of intermediate quality as it is based on few spectral features. On the basis of multiple lines of reasoning (see Section \ref{sec-pops} and Appendix \ref{sec-append}), we assume the velocity measurement is in error and this is a background galaxy. Thus, we identify a total sample of 38 Coma cluster galaxies with detected radio emission.

\subsection{Additional Sources}\label{sec-adds}

The catalog of MHM2008 was constructed by requiring a peak flux density greater than 4.5 times the local rms noise in order to fit and catalog a radio source. This means that faint extended sources with peaks below the 4.5$\sigma$ threshold will be missed. We therefore employed additional techniques to identify any low surface brightness radio sources. The radio mosaic images of Coma 1 and Coma 3 were convolved with a Gaussian to degrade their resolution from 4\farcs4 to each 7\farcs5 and 15\arcsec. We then searched for significant detections (5 times the local rms noise or greater) in these convolved images that were missed in the original radio source catalog and were coincident with SDSS optical sources having $r \leq 22$. This procedure identified one additional radio source, KUG1254+274 \citep[J125634.63+271339.4, with $cz = 7197$ \kms][]{castander2001}. This spiral galaxy was detected with a peak signal-to-noise of 5 in the 15\arcsec{} convolved image, and was resolved with a flux density of $1330 \pm 352$ $\mu$Jy. It was subsequently identified in the original 4\farcs4 resolution radio image, with a peak flux density of $70 \pm 28$ $\mu$Jy and an integrated flux density of $1160 \pm 485$ $\mu$Jy. No other faint resolved sources were identified.

\section{The Coma Cluster Radio Source Population}\label{sec-pops}

The radio-detected galaxies range from $r\approx12$ to $r\approx20$ (refer to Table \ref{tbl-mems}), with the majority within two magnitudes of $M^*$ \citep[$M^*_r \approx -21.5$, which translates to $r = 13.5$ for Coma;][]{blanton2001}. To maximize the efficiency of our spectroscopic observations using the MMT and Hectospec we generally did not observe sources that had redshifts listed in NED. Thus, the Hectospec observations were of fainter galaxies and hence the majority of the radio-detected Coma members have redshifts from other sources. In some cases this means we do not have their optical spectra and hence are not able to directly classify the galaxies as AGN or star-forming on the basis of absorption and emission features. We used a variety of data and techniques to assess the origin of the radio emission -- star formation or an active nucleus -- in the detected galaxies. These included examination of spectroscopic emission and absorption features (when optical spectra were available), best-fitting optical photometric profiles, optical and near-UV/optical colors, and prior classifications from the literature. The results are summarized in Table \ref{tbl-props}, where they follow the same order in increasing $r$ magnitude that was used in Table \ref{tbl-mems}. Specific comments on eight sources that proved slightly more difficult to classify are provided in Appendix \ref{sec-append}. The main properties we investigated to produce the AGN/star-forming classifications will now be described.

First, we have spectra for 27 objects through the SDSS DR6 \citep{adelman2008}, our Hectospec observations, and spectra collected for the NOAO Fundamental Plane Survey \citep[NFPS;][]{smith2004}. In addition to these sources, spectroscopic classifications for some of the other 11 radio-detected Coma members are indicated in the literature. The classifications based on the optical spectra are simplistic, and translate roughly to the spectroscopic classes used in the ``MORPHS'' work \citep{dressler1999,poggianti1999}. They are: 1) ``OSP'' for spectra consistent with an old stellar population and lacking emission lines. These correspond to the ``k'' classification of MORPHS. Radio emission from such galaxies is nearly always associated with an AGN \citep[e.g.,][]{miller2002,mauch2007}. 2) ``SF-burst'' for galaxies with strong emission lines indicative of high levels of current star formation. The continuum for this class of galaxies is often blue, as would be expected for a significant population of young massive stars. The relative strengths of their [{\scshape O~ii}] emission and H$\delta$ absorption lines would usually place them in the ``e(a)'' class used by MORPHS ([{\scshape O~ii}] detected in emission, EW(H$\delta$) absorption of greater than 4$\mbox{\AA}$), although the spectral resolution of both the SDSS and Hectospec data were sufficient to often identify H$\delta$ in emission within a broader absorption trough. Thus, at lower spectroscopic resolution the H$\delta$ from these galaxies would appear to be entirely in aborption. 3) ``SF-cont'' for lower levels of star formation. These galaxies had continua representative of older stellar populations, but the spectra also included weak emission lines indicative of star formation. They translate to the ``e(c),'' or continuous star formation class of MORPHS (EW([{\scshape O~ii}]) detected in emission but less than 40$\mbox{\AA}$, weak EW(H$\delta$) absorption of less than 4$\mbox{\AA}$). 4) ``Seyfert'' for an emission line galaxy with line diagnostics \citep{baldwin1981} showing photoionization from non-stellar radiation. Examples of each spectral type are depicted in Figure \ref{fig-specex}. The wavelength coverage of the NFPS spectra is about 4000$\mbox{\AA}$ to 6500$\mbox{\AA}$ and thus does not include [{\scshape O~ii}] or H$\alpha$. However, since they were obtained primarily for investigation of the fundamental plane they are almost exclusively elliptical galaxies and hence have OSP spectra. The two NFPS spectra for non-OSP galaxies have obvious Balmer emission lines and resulting classifications that are confirmed by other literature sources (KUG1258+278 and KUG1258+279A, with spectral classes of SF-cont and SF-burst, respectively). Of our 27 galaxies with available spectra, 15 are classified as OSP (and hence AGN), six as SF-burst, five as SF-cont, and one as Seyfert. A further six galaxies had starburst classifications indicated by NED.

Prior studies also provide strong expectations for the nature of radio-detected members based solely on optical photometry (i.e., sources for which we lack an optical spectrum but can rely upon a published redshift). Strong emission line AGN are rare in galaxy clusters \citep{dressler1985,cmiller2003}, and the majority of radio-detected cluster AGN are associated with elliptical galaxies that entirely lack optical emission lines or include only slight emission of the [{\scshape N~ii}] $\lambda 6584$ and [{\scshape S~ii}] $\lambda\lambda$(6716+6731) lines \citep{miller2002}. It is therefore generally safe to assume that radio-detected elliptical galaxies are AGN and non-ellipticals (spirals and irregulars) are star-forming, and the first avenue toward such segregation comes from the SDSS photometric data. The SDSS provides information on profile fits to galaxies and hence their basic morphology since elliptical galaxies are well fit by de~Vaucouleurs profiles whereas the disk nature of spirals are better fit by exponentials. Table \ref{tbl-props} indicates whether a given radio-detected galaxy was better fit by a de~Vaucouleurs profile (D) or an exponential (E), and hence whether its profile is suggestive of an elliptical or spiral morphology. We also performed a visual examination of the radio-detected galaxies using the SDSS ``Explore'' tool, and found that all galaxies that appeared to be ellipticals and S0s did have de~Vaucouleurs profiles. However, seven galaxies that had the appearance of spirals or irregulars were better fit by de~Vaucouleurs profiles than by exponentials. These were the larger cluster spirals where prominent bulges influence the fit (NGC4819, NGC4911, and NGC4853), some of the Markarian objects (MRK0055, MRK0056, and MRK0058), and GMP4582.

Photometric data also provide color information that can separate elliptical galaxies from spirals. We used $(u - g)$ as the measure of color since these filters straddle the $4000\mbox{\AA}$ break, and followed the procedure of \citet{lopezcruz2004} to fit the red sequence. The resulting fit was $(u - g) = 3.192 - 0.095r$ with a dispersion about this line of 0.060. Using this relation we defined the offset of a radio galaxy's color from the fitted red sequence as $\Delta (u - g) = (u - g)_{actual} - (u - g)_{RS} / 0.06$, where $(u - g)_{RS}$ was calculated from the measured $r$ magnitude of that galaxy. This provides a statistical measure of each galaxy's deviation from the red sequence, and a histogram of the results is shown in Figure \ref{fig-cmdhist}. The unshaded histogram is composed of all SDSS objects with $r \leq 17$ and within the survey area (ignoring possible spectroscopy to confirm or refute cluster membership), while the shaded portion represents only the radio-detected members of the Coma cluster. To grossly reveal the shift between bright cluster ellipticals and fainter cluster star-forming galaxies, we depict the radio galaxies with $r \leq 14.5$ as the darker portion of the shaded histogram. 

Near-ultraviolet (NUV) photometry is also useful in assessing recent star formation over $\sim10^9$ year timescales, and thus $(NUV - r)$ is a good indicator of the current to past-averaged SFR \citep{salim2005}. In fact, color-magnitude diagrams depicting $(NUV - r)$ and $M_r$ provide a cleaner separation of the red sequence and blue sequence \citep{wyder2007}. Furthermore, while optical color-magnitude diagrams are well described by two Gaussian fits -- one to the red sequence and one to the blue sequence -- $(NUV - r)$ vs.\ $M_r$ color-magnitude diagrams reveal an excess population between these sequences and believed to be transition objects, or galaxies in the process of moving onto the red sequence \citep{wyder2007,schiminovich2007}. In a deep {\it Galaxy Evolution Explorer} \citep[GALEX;][]{martin2005} study of the Coma 3 region, \citet{hammer2008} also found dwarf galaxies with low metallicity old stellar populations in this intermediate $(NUV - r)$ color region, but these objects are at fainter absolute magnitudes ($M_r \gtrsim -18$) than our radio-detected Coma galaxies. We have collected NUV photometry from the GALEX archive for the radio-detected Coma galaxies. A matching radius of 4\arcsec{} between the SDSS and GALEX positions was used, along with the photometry determined for elliptical Kron apertures \citep[i.e., the standard procedures for matching SDSS and GALEX sources and evaluating their colors; for example, see][]{bianchi2007}. The $r$ and $NUV$ photometry were corrected for Galactic extinction using the \citet{schlegel1998} maps, with the $NUV$ extinction determined as $A_{NUV}/E(B-V) = 8.741$ \citep{wyder2005}. For the optical magnitudes included in our study, $(NUV - r) \gtrsim 4.5$ are red sequence objects (AGN) and those with $(NUV - r) \lesssim 3.5$ are blue sequence objects (star-forming), while those intermediate are the transition objects (usually star-forming, depending on optical spectroscopy and other tests; see Appendix \ref{sec-append}). These $(NUV - r)$ ranges are the same as those used by \citet{hammer2008}. Figure \ref{fig-nuvmr} shows the $(NUV - r)$ vs. $M_r$ diagram constructed for the survey area. Points correspond to only those sources with GALEX photometry and redshifts consistent with Coma membership. Figure \ref{fig-nuvmr}  produces the same general impression as an color-magnitude diagram, although with a clearer separation between the radio AGN in massive ellipticals and the radio-detected star-forming galaxies. 

The availability of multi-filter photometry also presents the opportunity for fitting the spectral energy distributions of galaxies, and thereby providing additional details about their star formation histories and total stellar masses. We have used the \citet{blanton2007} K\_CORRECT software (v4\_1) for this purpose, which takes input redshifts and photometric data and determines the combination of model template spectra that best fits these data. The templates were based on numerous \citet{bruzual2003} models representing a range of metallicities, ages, and dust models, along with ionized gas models from \citet{kewley2001}. The star formation histories of the templates are thus known, and the best-fit combination of the templates for an individual galaxy thereby describes its star formation history and total stellar mass. We followed the K\_CORRECT recommendations regarding slight offsets to SDSS and GALEX magnitudes and errors, and included the derived stellar masses for all spectroscopically-confirmed Coma members in Table \ref{tbl-props}. We also inspected the individual synthetic spectra that K\_CORRECT determined as the best fits to the photometric data. These synthetic spectra provide an additional indication to whether a galaxy had active star formation or was better described as passive (i.e., lacking emission lines and fit by an old stellar population). Because the input photometric data represent the entire galaxy this can also provide a general notion of the location of any activity within that galaxy when compared with existing observed spectra; the SDSS and Hectospec fibers are 3\arcsec{} and 1\farcs5, respectively, and thus probe only the central kiloparsec or two.

Table \ref{tbl-props} summarizes the properties described above (classification from available nuclear spectrum, best-fit optical profile, $\Delta (u - g)$, $(NUV - r)$, and estimated stellar mass) and the resulting source categorizations as AGN or star-forming galaxies. In most cases the individual properties for a given galaxy produce a consistent view of its activity type, although eight sources with less certain classifications are discussed in the Appendix. To summarize, we placed equal numbers of galaxies in each of the AGN and star-forming classes. Nineteen galaxies were assigned an AGN classification, with two of these being somewhat uncertain: NGC4819 with red colors but a spiral morphology, and ARK395 with a Seyfert spectrum but likely including some radio emission associated with star formation. There was greater uncertainty in the star-forming class, as five of the 19 objects (NGC4853, IC3949, KUG1258+278, GMP4582, and MRK0055) proved challenging to classify on the basis of the available data. Finally, as previously indicated one of the radio galaxies with a reported velocity placing it within the Coma cluster (J12591536+2746052) was deemed a background galaxy. In all subsequent discussion, we will indicate where these less certain classifications affect our conclusions.

\section{The Coma Cluster Radio Luminosity Function}\label{sec-rlf}

In determining the RLF for Coma we follow the general procedure used by \citet{mobasher2003}. Lacking knowledge of the three-dimensional galaxy distribution we express our RLF in terms of surface density as the number of radio galaxies per area per ``magnitude.'' Here, magnitude is placed in quotes as the radio sources are binned by luminosity; a standard convention is to make these bins 0.4 dex in size, thereby paralleling the optical definition of a magnitude. Due to the fairly low number of sources we adopted larger bins of 0.6 dex for the radio souces. The faintest bin corresponds to sources with $20.0 \leq \log (L_{1.4}) < 20.6$, which translates to flux densities between 85 and 335 $\mu$Jy (equivalent SFRs from 0.06 to 0.23 \myr). The radio survey has a typical rms sensitivity of 28 $\mu$Jy and reaches down to 22 $\mu$Jy, meaning that the 5$\sigma$ detection threshold is always 110 $\mu$Jy ($\log (L_{1.4})$ = 20.12) or greater and thus fainter sources in this radio luminosity bin can not be detected over the full area. A correction for this narrower bin size was applied, and the individual radio counts within this bin were multiplied by factors to account for the reduced area coverage. For each galaxy this factor is the ratio of the full area to the area covered at the sensitivity needed to detect that galaxy at the 5$\sigma$ level.

RLFs are necessarily bivariate in the sense that redshift determination is usually based on optical spectroscopy, and thus the spectroscopic completeness factors into the derived RLF. All radio sources with optical counterparts having $r \lesssim 18$ have optically-determined redshifts, and hence we require no completeness correction to the radio counts below this optical magnitude limit. Based on Figure \ref{fig-specsamp}, applying an optical magnitude cut of $r \leq 18$ would remove only one potentially-confirmed cluster galaxy, J12591536+2746052 at $r \approx 20.2$. As discussed above and in Appendix \ref{sec-append} the velocity measurement for this galaxy is questionable and it is more likely to be a background object. Aside from this object, our spectroscopy data find no radio-detected galaxies belonging to Coma at magnitudes fainter than $r = 16.24$. Thus, methods which use the spectroscopic membership fraction (i.e., the number of radio-detected galaxies with spectroscopically-confirmed cluster velocities divided by the total number of radio-detected galaxies with measured velocities, as a function of magnitude) to correct for radio-detected optical sources lacking a velocity measurement would not produce any change in the RLF. We present the derived RLF in Figure \ref{fig-rlf}, noting that it is complete for all galaxies down to $r=18$ ($M_r = -17$) and is likely accurate to much fainter optical limits. Further discussion on possible radio-detected Coma members at fainter magnitudes will be provided in Section \ref{sec-discuss}. The error bars in Figure \ref{fig-rlf} are based entirely on Poisson statistics using the number of radio galaxies within each bin. 

We also divided the RLF into its respective contributions due to sources powered by an active nucleus and star formation. The fairly low number of Coma members produces relatively large error bars, although general trends are still evident. As expected, AGN are the sole contributor to the RLF at the higher luminosities ($L_{1.4GHz} \gtrsim 2 \times 10^{22}$ \whz). Star forming galaxies then become the more numerous population, although surprisingly below about $10^{21}$ \whz{} the AGN again become the dominant population of radio-emitting galaxies. Figure \ref{fig-rlf} also includes representations of the fitted functional forms of the AGN and star-forming RLFs, as parametrized by \citet{condon2002} for local field samples. The relative normalization of these separate RLFs was taken from \citet{lin2007}, and can be seen to do a reasonable job of matching the total RLF and the brighter end portions of the AGN and star-forming RLFs.

\section{Discussion}\label{sec-discuss}

Our motivation for this work was to explore the RLF, in particular for fainter objects (both in terms of radio luminosity and optical magnitude). We will first turn our attention to the relative populations of AGN and star-forming galaxies at radio luminosities below about $10^{21}$ \whz, a regime that has hitherto not been explored for galaxy clusters. Next, we will address the radio-identified star-forming galaxy population and possible clues to cluster galaxy evolution that such objects provide. Finally, we will investigate the AGN uncovered by our radio data.

\subsection{The Decline of Star Formation Relative to AGN at Low Radio Luminosities}\label{sec-declinesf}

The dearth of star-forming galaxies at lower luminosities (Figure \ref{fig-rlf}) is unexpected. Deep RLFs derived for large samples of nearby star-forming galaxies (and hence dominated by non-cluster, field environments) either flatten or continue to rise slowly at fainter radio luminosities \citep{condon2002,mauch2007}. The RLFs for AGN continue to rise, although star-forming galaxies are still more numerous at these faint radio luminosities. For example, at $10^{20}$ \whz{} these field RLFs indicate about ten star-forming galaxies per AGN \citep[refer to Figures 9 and 12 of][]{condon2002,mauch2007}. The general shapes of the respective field RLFs suggest that at very faint radio luminosities (below $10^{20}$ \whz) AGN may once again dominate the radio source population, although radio luminosities this faint have not been probed by field RLFs to date. The \citet{condon2002} RLF for star-forming galaxies has a faint-end limit of about $10^{19}$ \whz, while that of AGN reaches to about $10^{20}$ \whz. The limits for the \citet{mauch2007} RLFs are similar, but not quite as deep. In our Coma RLF, we are finding that the star-forming galaxies are the dominant radio-emitting population only over a small range in radio luminosity, roughly from $10^{21}$ to $10^{22}$ \whz. Below about $10^{21}$ \whz, AGN are more numerous than star-forming galaxies for our Coma cluster data. Previous RLFs derived for clusters, with the exception of the Virgo work by GB99, only extend down to radio luminosities of a few $\times 10^{21}$ \whz \citep{miller2002,lin2007}. Thus, there is no real precedent to this intriguing result and it is well worth additional scrutiny.

The obvious question to ask is: are we somehow missing the lower radio luminosity star-forming galaxies? One consideration is that our interferometric data might fail to detect low levels of radio emission spread over the extents of weaker star-forming galaxies, but still successfully detect compact radio emission arising from AGN. The synthesized beam of the radio observations was 4\farcs4, which equates to about 2 kpc at the distance of Coma. Although this seems a good match to the expectation that the supernova-accelerated particles emit synchrotron radiation over roughly kpc scales \citep{condon1992}, such diffuse emission will not be detected by all VLA baselines and hence the sensitivity to such emission will be worse than that for sources unresolved on all baselines. Figure \ref{fig-ctsbyres} demonstrates this differing sensitivity to resolved and unresolved sources by providing histograms of the flux densities of all radio sources within the survey area broken down by whether they are resolved. The peak in the counts for unresolved sources occurs about 100 $\mu$Jy fainter than the peak in the counts for resolved sources. If star-forming galaxies are associated with resolved radio sources whereas AGN are more likely to be associated with unresolved sources, the lack of star-forming galaxies at the faint end of the RLF might simply be a selection effect. Indeed, we find that the radio emission from all of our star-forming galaxies was resolved by the 4\farcs4 synthesized beam of the radio observations, while seven of the 19 AGN were unresolved. We note that the first bin in the RLF of Figure \ref{fig-rlf} extends from 85 $\mu$Jy to 335 $\mu$Jy (equivalent to SFR from 0.06 \myr{}  to 0.23 \myr) and the peak in the counts of the resolved sources in Figure \ref{fig-ctsbyres} occurs near the upper limit of this bin at somewhere between 250 and 300 $\mu$Jy. The decline in the star-forming population is evident in the second bin of our RLF (335 $\mu$Jy to 1.334 mJy), providing additional support to the argument that the decline in star-forming galaxies relative to AGN is real.

Our examination of convolved radio images (see Section \ref{sec-adds}) also provides some insight into this issue. Fainter resolved sources can miss the cutoff of a peak flux density greater than 4.5 times the local rms sensitivity, but still include multiple contiguous resolution elements at lower significance. Convolving with a Gaussian of an appropriate size effectively collects these lower significance detections and can produce a single higher significance detection associated with coarser resolution. This procedure allowed us to identify KUG1254+274 as a cluster radio source. Of course, this is not ``magic.'' The rms noise of the convolved image is higher than the rms noise of the original by the square root of the ratio of the respective beam areas, which for circular beams reduces to the ratio of their diameters. Thus, even for our 7\farcs5 resolution convolved image the typical rms noise was near 50 $\mu$Jy per beam -- sufficient only to detect the very brightest sources in the first bin of the RLF shown in Figure \ref{fig-rlf}. 

Similarly, we can resort to stacking of sources in an attempt to reveal faint radio emission associated with star-forming galaxies. We used the fit to the color-magnitude diagram to identify all blue galaxies within the survey area, where blue was defined as a greater than 2$\sigma$ separation from the fitted red sequence. We then correlated this list with the spectroscopic database of Marzke et al., thus producing a listing of Coma galaxies with blue colors yet undetected by the radio observations. For each of the resulting 45 galaxies down to $r = 18$ ($M_r = -17$), a 2\arcmin{} square cutout of the radio mosaic centered at the optical position was made. These cutouts were then stacked to examine the average radio properties of the blue population. \citet{white2007} discussed the relative merits of using the mean or median in stacking of sources from the FIRST survey, and opted to use the median to remove the detrimental effects of outliers (individual images in the stack with unusually poor noise characteristics, real sources outside the center position, etc.). We chose to use the mean but included rejection of the minimum and maximum value for each pixel in the stack. The stacked image had an rms of under 5 $\mu$Jy per beam, yet no radio emission associated with the overall blue galaxy population was identified. Subsets of the 45 galaxies determined by optical magnitude (e.g., $r < 16$, $16 \leq r < 17$, $17 \leq r < 18$) also did not produce any significant detections.

We also examined all the SDSS spectra for objects within the survey area in order to identify any Coma cluster emission line galaxies that were not among our radio detections. Because the fibers used for collection of spectra in the SDSS are 3\arcsec{} (about 1.4 kpc), this selection is also insensitive to star formation spread over galaxy disks yet it still provides an additional avenue toward source characterization. We found eight emission line galaxies in this manner, all of which had blue colors and hence were included in the stacking analysis. Seven of the eight emission line galaxies had spectra indicative of star formation with the eighth having a Seyfert spectrum. Four of the seven star-forming galaxies were similar to GMP4582 (one of our radio detections) with spectra dominated by old stellar populations but including weak emission lines consistent with low levels of star formation. Their photometric properties were also consistent with those of GMP4582, representative of galaxies with stellar masses of a few times $10^9$ M$_\odot$ and lying just off the red sequence. The other three were starbursts in fainter dwarf galaxies ($M_r \approx -18$). Two of the starbursts and one of the more quiescent star-forming galaxies were found to be associated with low significance radio sources (between 2$\sigma$ and 3$\sigma$), while the remaining galaxies lacked even a $2\sigma$ peak flux density within 5\arcsec{} of their optical positions. Repeating the stacking analyis for just these eight emission line galaxies did yield a weak radio detection, with a peak flux density of $43 \pm 11$ $\mu$Jy (equivalent to SFR = 0.03 \myr). This is roughly equivalent to the expected radio flux density determined from the GALEX FUV data under the assumption that the emission arises from active star formation; conversion of the FUV flux densities to SFRs using the relationship of \citet{kennicutt1998} produces SFRs ranging from 0.005 \myr{} to 0.18 \myr. This suggests that our radio mosaic is not terribly biased against detecting faint star-forming galaxies, as they appear to be present but below the formal detection limit.
% Actual average SFR of EL non-detects is 0.06, about twice radio value

The decline of cluster star-forming galaxies also seems to be countered by an increase in AGN. The total Coma RLF (Figure \ref{fig-rlf}) is reasonably well fit by the sum of the \citet{condon2002} AGN and star-forming RLFs, with overall normalization as determined by \citet{lin2007} for a large sample (nearly 600 objects) of nearby X-ray selected clusters. \citet{lin2007} used $K$-band data and a statistical approach to separate cluster members from background galaxies, and hence were unable to classify individual galaxies as star-forming or AGN. They could, however, fit the total cluster RLF as the sum of the AGN and star-forming populations with the ``dip'' in the total RLF thereby constraining the inferred contributions from the AGN and star-forming galaxy RLFs. The \citet{lin2007} RLF only reached radio luminosities of a few $\times 10^{21}$ \whz, which is just above the location in the Coma RLF where star formation appears to decline at the expense of an increased AGN population. However, the notion that the overall cluster RLF as found by \citet{lin2007} produces a good match to the Coma RLF suggests that the decline in star formation at lower radio luminosities is offset by an increase in AGN. One explanation for this is simply incorrect source classification, such that weak star-forming galaxies are classified as AGN. Yet we find that our faint radio-luminosity AGN show little evidence for current star formation: they have optical spectra of solely absorption lines consistent with a dominant stellar population of older stars, their optical colors and profiles are those of elliptical galaxies (with the exception of NGC4819, which has a red color and de~Vaucouleurs profile yet has an early-type spiral morphology), and they show no evidence for current or recent star formation in their GALEX UV photometry. 
 
To summarize, the apparent lack of lower SFR galaxies observed among the radio-detected members of the Coma cluster appears to be real but we can not rule out selection effects caused by the surface brightness sensitivity of our interferometric radio data. The sensitivity of the radio data is insufficient to detect the majority of sources within the first bin of the RLF of Figure \ref{fig-rlf} should they be resolved, but attempts to reveal the presence of such a population through stacking produced a detection only for a small subset of galaxies (those having optical emission lines). The stacked image for these emission-line identified galaxies had a radio luminosity of about $5 \times 10^{19}$ \whz, below the first bin of our RLF. The equivalent SFR is consistent with that determined from the FUV emission of the galaxies, suggesting that resolution effects were not important: the radio emission associated with star formation in these galaxies is present in the data, but lies below the formal detection limit. Furthermore, the apparent downturn in the RLF for star-forming galaxies arises in the second bin of the RLF where our data are complete, with this bin spanning from $\log L_{1.4GHz} = 20.6$ to 21.2 (SFRs from 0.23 \myr{} to 0.93 \myr). This supports the notion that star-forming galaxies at fainter radio luminosities are becoming increasingly rare. One further important point is that the physical size of the fainter dwarf starburst galaxies selected via their emission line spectra (Petrosian radii of several arcseconds) is approaching the resolution of the radio images. This argues that any smaller dwarf starbursts would be unresolved by our radio data and hence more easily detected should their SFRs be large enough. A final piece of evidence that suggests we are not missing star-forming galaxies at fainter radio luminosities is that their apparent decline is countered by an increase in AGN; the total RLF is consistent with expectations, it is just the relative contributions of AGN and star-forming galaxies that differ. We conclude this section by noting that the most direct test will be to obtain additional radio continuum observations of the Coma cluster with increased sensitivity to lower surface brightness extended radio emission, as well as producing radio and optical data sets for other nearby clusters down to comparable sensitivity limits.

\subsection{The Prevalence of Starbursts}\label{sec-bursts}

Another interesting point regarding the detected star-forming galaxies is that they include a high proportion of starbursts as opposed to more quiescent galaxies with moderate SFR. Between spectroscopic classifications from the literature and those determined from available SDSS and Hectospec spectroscopy, 12 of the 19 identified star-forming galaxies are starbursts. We use the radio-derived SFRs to further elucidate this point in Figure \ref{fig-ssfr}. The specific SFR (SSFR) is defined as the current star formation rate per unit stellar mass, so for our identified star-forming galaxies we convert the radio luminosities to SFRs using the relationship of \citet{yun2001} and divide this by the stellar masses obtained by fitting the photometric data with K\_CORRECT \citep{blanton2007}. The starburst classifications made on the basis of optical spectra (Table \ref{tbl-props}) are largely confirmed, with 11 of the 12 clustered at SSFRs around $10^{-9}$ yr$^{-1}$ in Figure \ref{fig-ssfr}. For comparison, the average SSFR for UV and FIR selected samples of local star-forming galaxies with stellar masses around $3 \times 10^9$ M$_\odot$ (a mass typical of our sample) is about $2 \times 10^{-10}$  yr$^{-1}$ \citep{buat2007}, and the Coma radio-detected starbursts lie near the putative upper limit to SSFR as a function of mass \citep[][; indicated by the dashed line in Figure \ref{fig-ssfr}]{feulner2006}. This indicates that these Coma galaxies are forming stars as vigorously as any observed galaxies of the same stellar mass. The only spectroscopically-identified starburst missing among the grouping of starbursts in Figure \ref{fig-ssfr} is IC3913 with a slightly lower SSFR.

Several factors might affect our radio-derived SFR estimates. First, there is the potential that the radio data are missing flux on account of the star-forming galaxies being resolved. This would underestimate the SFRs and subsequently SSFRs. Comparison with the NVSS indicates that this is not a problem, with the MHM2008 and NVSS fluxes consistent for the 15 objects in common (these include AGN as well as star-forming galaxies). Only IC4040 had an NVSS flux density that was marginally higher than the MHM2008 value we used ($21.1 \pm 1.1$ mJy vs. $18.31 \pm 0.16$ mJy). Running counter to this is the possibility that the radio overpredicts the SFRs due to the possible inclusion of an AGN or through some environmentally-induced increase in radio luminosity arising from star formation. The radio luminosities of many of the Coma star-forming galaxies are about twice what would be expected based on their far-infrared fluxes, which could be due either to an AGN contribution to the net radio flux or interaction with the intracluster medium \citep[e.g., increasing the synchrotron power by compressing magnetic fields via ram pressure or thermal compression][]{miller2001b,reddy2004}. High resolution radio observations designed to assess any AGN contribution to the net radio flux in a sample of comparable galaxies suggested low contamination by AGN \citep{miller2001b}. These included two of the current sample of Coma galaxies, NGC4911 and IC4040. NGC4911 potentially does include an AGN as revealed by a compact ($<0\farcs12 \times <0\farcs12$, or under 60 pc) source with an 8.46~GHz flux density of 0.73 mJy. NGC4911 is resolved in our 1.4~GHz data with a flux density of 19.09 mJy and radio contours that trace its disk (see Figure 5 of MHM2008). Should the unresolved 8.46~GHz detection correspond to an AGN, a flat spectrum would mean less than 4\% of the total 1.4~GHz flux density is attributable to the AGN. We note that the peak flux density in our 4\farcs4 resolution 1.4~GHz data is 0.75 mJy, effectively ruling out a steep spectral index and hence higher 1.4~GHz flux density contribution for any AGN. IC4040 was not detected in the 8.46~GHz observations, down to an rms sensitivity of 50 $\mu$Jy per 0\farcs2 beam. Finally, we find that the SFRs derived using the radio data are consistent with those determined from the GALEX {\it FUV} magnitudes assuming $A_{FUV}$ values ranging from about 0.7 to 3.6 magnitudes, with the majority between 1 and 2 magnitudes. These extinctions are typical of star-forming galaxies selected via GALEX and SDSS emission line samples \citep[e.g.,][]{salim2007}.

In conjunction with the lack of star-forming galaxies detected at fainter radio luminosities, this is suggestive of the same general evolutionary scenario envisaged for intermediate redshift clusters studied by the MORPHS group \citep{dressler1999,poggianti1999}. As star-forming galaxies are accreted by clusters, some mechanism promotes strong bursts of star formation which are then rapidly extinguished, leading to increased populations of post-starburst galaxies. \citet[][hereafter P2004]{poggianti2004} investigated the post-starburst population of the Coma cluster and compared it to those of the more distant clusters of the MORPHS sample. They found that post-starbursts were absent among brighter Coma members ($M_V \leq -20$) whereas such galaxies are numerous in distant clusters. However, Coma does contain a substantial population of post-starburst galaxies among its fainter population with around 15\% of its galaxies with $M_B \gtrsim -17.3$ exhibiting such spectra. This is consistent with ``downsizing.'' Earlier studies had also identified Coma galaxies with evidence for recent starbursts \citep[e.g.,][and subsequent papers by these authors]{caldwell1993} and suggested a connection with the dynamical environment of Coma. P2004 refined this link, noting that the post-starbursts with bluer optical colors correlated well with X-ray substructure whereas those post-starbursts with redder colors traced the overall galaxy distribution more directly. The colors and Balmer absorption-line strengths of the blue post-starbursts indicate recent truncation of starburst episodes (within a few hundred million years) rather than more gradual processes such as ``strangulation'' (the extinguishing of normal star formation as the gas reservoirs of galaxies are removed) which can explain the colors and absorption line strengths of the red post-starbursts. In addition, the velocity dispersion of the blue post-starbursts was significantly higher than those corresponding to quiescent galaxies and red (i.e., older) k+a's.

It is illustrative to compare our radio-detected starbursts with the post-starbursts of P2004. In Figure \ref{fig-dist} we plot the distributions of the post-starbursts along with our star-forming galaxies. As indicated in P2004, the overall distribution appears well mixed. However, P2004 found that the blue post-starbursts (open pentagons in Figure \ref{fig-dist}) matched with a pair of X-ray substructures, one running approximately north-south and coincident with the western/eastern edge of Coma 1/Coma 3, and a smaller one aligned roughly east-west and jutting into Coma 1 from its eastern edge at a declination of about 27\fdg75. The radio-identified starbursts also appear to be arranged along this north-south axis of X-ray substructure, with four starbursts located along the eastern edge of Coma 3 and two along the western edge of Coma 1. In addition, three more starbursts in Coma 1 encircle a grouping of three blue post-starbursts (i.e., the objects which best trace the east-west X-ray substructure). We suggest a common origin for these two populations, in the sense that they have similar dynamical histories but with one class being viewed shortly before the termination of its starburst episode and the other being viewed shortly after. The optical magnitudes of the P2004 blue post-starbursts are typically around $M_r = -18$, in line with the expected fading of our starbursts should their star formation terminate.

To quantitatively test the possible correpondence between radio-identified starbursts and the P2004 blue post-starbursts, we performed an analysis on pair separations. Four classes of object were used: 1) radio-identified starburst galaxies, 2) radio-identified star-forming galaxies with more gradual star formation histories, 3) blue post-starbursts of P2004, and 4) red post-starbursts of P2004, whose colors and Balmer absorption line strengths indicate a longer time since the ending of their star-formation ($\sim$1 Gyr). The separation to the nearest neighbor from each of the four classes was determined for every galaxy, and the mean separation for each class pairing was tablulated (Table \ref{tbl-pairs}). For example, the average distance to the nearest blue post-starburst galaxy from a radio-identified starburst was 670\arcsec{} (note that the converse, the average distance to the nearest radio-identified starburst from a blue post-starburst, is 558\arcsec{} since there 12 radio-identified starbursts but only 11 blue post-starbursts). 

We then performed Monte Carlo (MC) simulations by shuffling the positions of the galaxies, recomputing the mean pair separations, and noting when these means were less than those determined for the real data. We used two separate shuffling procedures and performed 1000 simulations for each. The first maintained the radial profiles of the four galaxy classes by holding their cluster-centric distances fixed (based on the adopted center position for the Coma 1 field) and performing the shuffle in azimuth, with the restriction that the positions in the shuffled data had to be within the Coma 1 and Coma 3 boundaries. The second shuffling procedure was entirely random in RA and Dec, with the only restriction being that galaxies initially within Coma 1 (Coma 3) remained within Coma 1 (Coma 3) after their positions had been randomized. In no case were the mean separations from the real data significantly less than those in the MC data (Table \ref{tbl-mcpairs}). There is possibly a slight tendency for the red post-starbursts to be located near the radio starbursts, with 14.1\% and 9.4\% of the simulations having lower mean separations than the actual data for the cases where the radial distance was fixed and where the positions were fully randomized, respectively. This seems to be more related to the real distribution of the red post-starbursts than to any physical pairing of the two types, as the converse (the mean separation from a radio starburst to the nearest red post-starburst) is not at all significant. Inspection of Figure \ref{fig-dist} shows that the red post-starbursts within Coma 1 are fairly evenly distributed in a ring roughly 20\arcmin{} in diameter and centered on the cluster center. It is also interesting to note that the individual types do not appear to be clustered to themselves (elements along the diagonal of Table \ref{tbl-mcpairs}). This would suggest large-scale structures as the origin of the various populations instead of the accretion of small groups of galaxies. In summary, the radio-identified starbursts appear to trace the same X-ray structures as the blue post-starbursts of P2004, yet we find no conclusive evidence of this connection when analyzing pair separations.

This qualitative correspondence between radio-identified starbursts and X-ray structures allows for some speculation on the more influential environmental factors. Numerous models predict the triggering of starbursts in infalling groups, including via tidal interactions \citep{bekki1999,gnedin2003} and via interaction with the intracluster gas of the cluster \citep[e.g.,][]{dressler1983}. Effects of this latter variety may be augmented by large-scale merger events, as shock fronts in intracluster gas may be produced by cluster mergers and individual galaxies might experience strong spikes in ram pressure induced star formation as they cross these fronts \citep{roettiger1996}. Thus, alignments with X-ray features as suggested by P2004 and potentially traced by our radio-detected starbursts might be expected. However, we conclude this discussion by emphasizing that our survey has specifically selected its covered area on the basis of previously-identified structure. More complete analyses will require deep radio imaging and optical spectroscopy over the entire cluster.

\subsection{AGN in Cluster Elliptical Galaxies}\label{sec-ellips}

As noted in MHM2008 and depicted in Figure \ref{fig-cmdhist}, the detection fraction for red sequence galaxies rises for $M_r \leq -20.5$. Of the 43 red sequence galaxies satisfying this optical magnitude cut, 18 are detected by the radio observations. In fact, all of the AGN in the RLF of Figure \ref{fig-rlf} are brighter than this magnitude limit, with the only AGN not lying on the red sequence being the Seyfert 2 galaxy ARK395 ($M_r = -20.6$, $L_{1.4GHz} = 2.2 \times 10^{21}$ \whz). We further investigated the red sequence objects that had no associated radio source through stacking. All 25 of the non-detected red sequence galaxies with $M_r \leq -20.5$ had optical spectroscopy providing velocity information, with 24 of them being cluster members. Stacking these 24 sources, we obtained a detection corresponding to an unresolved source with a flux density of $25 \pm 6$ $\mu$Jy ($L_{1.4GHz} = 3 \times 10^{19}$ \whz). Therefore, some of these optically-bright red sequence galaxies do contain an AGN below the formal radio detection limit of our survey. Furthermore, it is plausible that all massive elliptical galaxies could contain an AGN detectable by radio observations, provided that the radio data are sufficiently deep. \citet{hodge2008} come to a similar conclusion based on stacking of FIRST data \citep{becker1995} for luminous red galaxies from the SDSS. The trend might continue to fainter elliptical galaxies, although a stacking analysis on all cluster-member red sequence galaxies with $14.5 < r \leq 16.0$ failed to produce a detection down to an rms sensitivity of 3.1 $\mu$Jy per beam in the stacked image.

Although we continue to identify AGN as we reach lower radio luminosities, we find no evidence for a population of optically-faint Coma dwarf galaxies hosting relatively powerful radio sources as reported by GB99 for the Virgo cluster. The GB99 study used NVSS radio data, implying a radio luminosity limit nearly identical to our limit for the Coma cluster. Thus, should such objects exist in Coma the challenge is not in detecting their radio emission but in obtaining the optical spectroscopy to assess their membership in Coma. Incomplete optical spectroscopy does not allow us to rule out such objects entirely, yet our measured redshifts for a significant fraction of the radio-detected objects among fainter galaxies does allow us to place limits on their possible existence. For example, if there were only one such galaxy in our survey area and it had $18 \leq r < 21$ ($M_r < -14$, the same approximate depth as GB99), we would collect a spectrum for it most of the time (73\%) under the assumption that our spectroscopic sampling was random. This is based on there being 133 radio-detected galaxies within this magnitude range, with spectroscopic redshifts existing for 97 of them (assuming the reported value for J12591536+2746052 is in error, and hence no redshift exists for this galaxy; see Appendix \ref{sec-append}). The spectroscopic sampling over this magnitude range is nearly constant, meaning we could perform the same calculation but restrict the magnitude range to $20 \leq r < 21$ and arrive at the same general result (42 total galaxies, of which 29 have spectroscopic redshifts, for a 69\% chance of observing such a galaxy). Considering GB99 identified about ten Virgo cluster dwarf ellipticals hosting radio sources and within this approximate optical magnitude range (Table 1 and Figure 1b of GB99, for $-13 \leq M_B < -16$), our lack of a single detected analog in Coma is a strong result. 

Although Virgo and Coma represent different environments and our study is restricted to only a portion of the full Coma cluster, we suggest that the GB99 finding was biased by false radio-optical associations and incomplete spectroscopic sampling. The NVSS has a 45\arcsec{} beam making false superpositions more likely, and although GB99 did include some redshift information, the majority of faint sources lacked spectroscopic measurements and could easily be background radio-emitting elliptical galaxies. To further test this, we re-examined the dwarf ellipticals from Table 1 of GB99. Using NED to remove objects with non-Virgo redshifts, we were left with six objects having photographic magnitudes fainter than 16 ($M_B \gtrsim -15$). Five of these lacked spectra but had SDSS photometric redshifts ranging from 0.34 to 0.82. The lone confirmed Virgo member (VCC \#1606) had a separation between its SDSS optical position and its nearest NVSS counterpart (an unresolved source with a flux density of 4.6 mJy) of 29\arcsec. 

% There are about 50 sources per square degree in the NVSS \citep{condon1998}, which indicates 
% that the probability that VCC \#1606 is unrelated to the radio source is only about 1\%. 
% NVSS 12:35:16.65 +12:14:12.0  unresolved, 4.6 mJy
% SDSS 12:35:14.69 +12:14:14.7

\section{Conclusions}

We have presented a deep RLF derived for the core and outskirts of the Coma cluster. This RLF reaches radio luminosities of just over $10^{20}$ \whz, about an order of magnitude deeper than prior cluster RLFs. Extensive optical spectroscopy makes our RLF complete for all galaxies down to optical magnitudes of $M_r = -17$. We identify no cluster radio galaxies fainter than this optical magnitude despite excellent spectroscopic completeness down to $r \approx 22$ ($M_r \lesssim -13$ should such objects lie within the Coma cluster). The 38 radio-detected Coma members are evenly split between those powered by AGN and those powered by star formation.

We identify very few star-forming galaxies at radio luminosities fainter than $\log (L_{1.4GHz}) \approx 10^{21}$ \whz. This is in contrast with local field RLFs which show that the dominant contributor to the RLF at these radio luminosities and fainter continues to be star-forming galaxies. We caution that this finding could be partially caused by the selection effect of the 4\farcs4 resolution of our data and the extended nature of radio emission associated with star formation, and note that this is a strong motivation for future cluster observations at a range of resolutions. However, the total decline in star-forming galaxies at the lower luminosities of the RLF is offset by an increase in AGN associated with massive elliptical galaxies, such that the total RLF is approximately consistent with those derived for samples of field galaxies.

The radio-detected star-forming galaxies include a large fraction of starbursts, with high specific star formation rates. The optical magnitudes of these galaxies are typically in the range of $-19 \geq M_r \geq -20$ with star formation rates of a few \myr. The presence of these galaxies and the absence of star-forming galaxies with lower star formation rates suggests that environmental processes within the Coma cluster lead to starbursts followed by rapid cessation of star formation. The magnitudes and distribution of the radio-identified starbursts suggest a connection with the blue post-starburst population described in \citet{poggianti2004}, although the two populations do not show any evidence for lower pair separations than expected based on random distributions of galaxies. 

Optically bright elliptical galaxies are frequent radio detections and appear to be a significant contributor to the galaxy cluster RLF at all radio luminosities. Stacking analysis of non-radio detected cluster ellipticals with $M_r \leq -20.5$ reveals an average radio luminosity of $\log (L_{1.4GHz}) \approx 3 \times 10^{19}$ \whz. However, contrary to the findings of \citet{gavazzi1999} for the Virgo cluster, we find no evidence for a population of radio-emitting strong AGN in optically faint ($M_r \approx -14$) cluster dwarf galaxies. Although incomplete optical spectroscopy limits this conclusion, we note that should even one such object exist within the Coma cluster our spectroscopic sampling indicates a better than 70\% probability that it would have been identified.

\acknowledgments
NAM gratefully acknowledges the support of a Jansky Fellowship of the NRAO, held during the period when the majority of this work was performed. We thank the anonymous referee for comments that helped clarify the text, and Dave Carter and the members of the HST/ACS Coma Treasury program for motivating this work.

Funding for the Sloan Digital Sky Survey (SDSS) has been provided by the Alfred P. Sloan Foundation, the Participating Institutions, the National Aeronautics and Space Administration, the National Science Foundation, the U.S. Department of Energy, the Japanese Monbukagakusho, and the Max Planck Society. The SDSS Web site is http://www.sdss.org/.

The SDSS is managed by the Astrophysical Research Consortium (ARC) for the Participating Institutions. The Participating Institutions are The University of Chicago, Fermilab, the Institute for Advanced Study, the Japan Participation Group, The Johns Hopkins University, the Korean Scientist Group, Los Alamos National Laboratory, the Max-Planck-Institute for Astronomy (MPIA), the Max-Planck-Institute for Astrophysics (MPA), New Mexico State University, University of Pittsburgh, University of Portsmouth, Princeton University, the United States Naval Observatory, and the University of Washington.

This research has made use of the NASA/IPAC Extragalactic Database (NED) which is operated by the Jet Propulsion Laboratory, California Institute of Technology, under contract with the National Aeronautics and Space Administration.

\appendix

\section{Notes on AGN/SF Classifications for Less Certain Sources}\label{sec-append}

\noindent
{\bf NGC4819} This galaxy has a relatively red optical color with $\Delta (u - g) = -2.45$ and is better fit by a de~Vaucouleurs profile. Consistent with these factors is its $(NUV - r) = 5.07$, which is near the red sequence but potentially slightly offset to the blue (see Figure \ref{fig-nuvmr}). Its morphological classification in NED is (R')SAB(r)a, and visually there is the suggestion of slightly blue patches of its disk. Its synthesized spectrum from K\_CORRECT is primarily that of a passive galaxy, although with weak emission of [{\scshape O~ii}] and H$\alpha$+[{\scshape N~ii}]. Prior studies have indicated that it is highly {\scshape H~i} deficient and only marginally detected in H$\alpha$+[{\scshape N~ii}] \citep[e.g.,][]{gavazzi2002}, consistent with its red color and fitted spectrum.  The radio emission is compact and coincident with the nucleus and not the disk, and this along with its red colors leads us to an AGN classification. Alternatively, it might have very low residual star formation in its nucleus (0.2 \myr) prior to its landing on the red sequence.

\noindent
{\bf NGC4853} The blue portion of this galaxy's optical spectrum is that of a post-starburst galaxy, lacking any [{\scshape O~ii}] emission and having fairly strong Balmer absorption. However, it has weak emission of H$\alpha$ and [{\scshape N~ii}] with line ratios (corrected for underlying stellar absorption) indicative of star formation. It has a substantial bulge and is better fit by a de~Vaucouleurs profile and its $(NUV - r) = 3.70$ suggests that it is a transition object. The synthesized spectrum that best fits the overall photometric data is that of a star-forming galaxy. The far-infrared (FIR) and radio fluxes for star-forming galaxies are tightly correlated \citep[e.g.,][]{condon1992,yun2001}, and NGC4853 lies near this correlation although it is somewhat overluminous in the FIR \citep{miller2001a}. We place it tentatively among the star-forming galaxies, although we note that an AGN contribution to the total radio luminosity can not be ruled out.

%Interestingly, it is also an X-ray detection with $L{0.5-2.0keV} = 2.2\pm0.3 \times 10^{40}$ ergs~s$^{-1}$ \citep{finoguenov2004}. This is significantly above the reputed correlation between X-ray and radio luminosity for star-forming galaxies \citep[e.g.,][]{ranalli2003}. We thus place NGC4853 among the AGN, although note that its radio luminosity likely includes some contribution from star formation.

% Xray SFR is 2.2E-40 x L0.5-2.0keV; that is probably for stars more massive than 5. 
%  IC4040   Xray 16.1e-15 flux -> 1.9e40 erg/s -> 4.2 Msun/yr; radio 12.8... yes, multiply xray SFRs
%  from Ranalli relation by 5.5.
%  NGC4853 Xray predicts 4.9 27.2 Msun/yr (second has 5.5 factor for IMF); radio is 1.8
%  IC3949 Xray predicts 3.2 17.4 Msun/yr (full IMF); radio is 1.8
%  NGC4858 Xray predicts 2.5 13.5 Msun/yr (full IMF), radio is 5.9
%  MRK55  Xray predicts 1.9 10.4 Msun/yr, radio is 1.5

\noindent
{\bf IC3949} This galaxy has an exponential profile and blue optical and $NUV$-optical colors indicative of star formation. However, its nuclear spectrum is mainly that of an older stellar population but with weak H$\alpha$ and [{\scshape N~ii}] emission lines. The equivalent width of the H$\alpha$ line is less than that of the [{\scshape N~ii}], suggestive of an active nucleus. Correcting the H$\alpha$ equivalent width for underlying stellar absorption, the [{\scshape N~ii}]/H$\alpha$ ratio is very close to 0.6, the value frequently used as the division between star-forming galaxies and AGN. The synthesized spectrum that best fits the photometry is that of a starburst. We interpret this galaxy as having its radio emission dominated by star formation in its disk with a more quiescent central bulge; visual inspection of the SDSS color image backs this intepretation.

\begin{figure}
\figurenum{6}
\epsscale{0.7}
\rotatebox{270}{\plotone{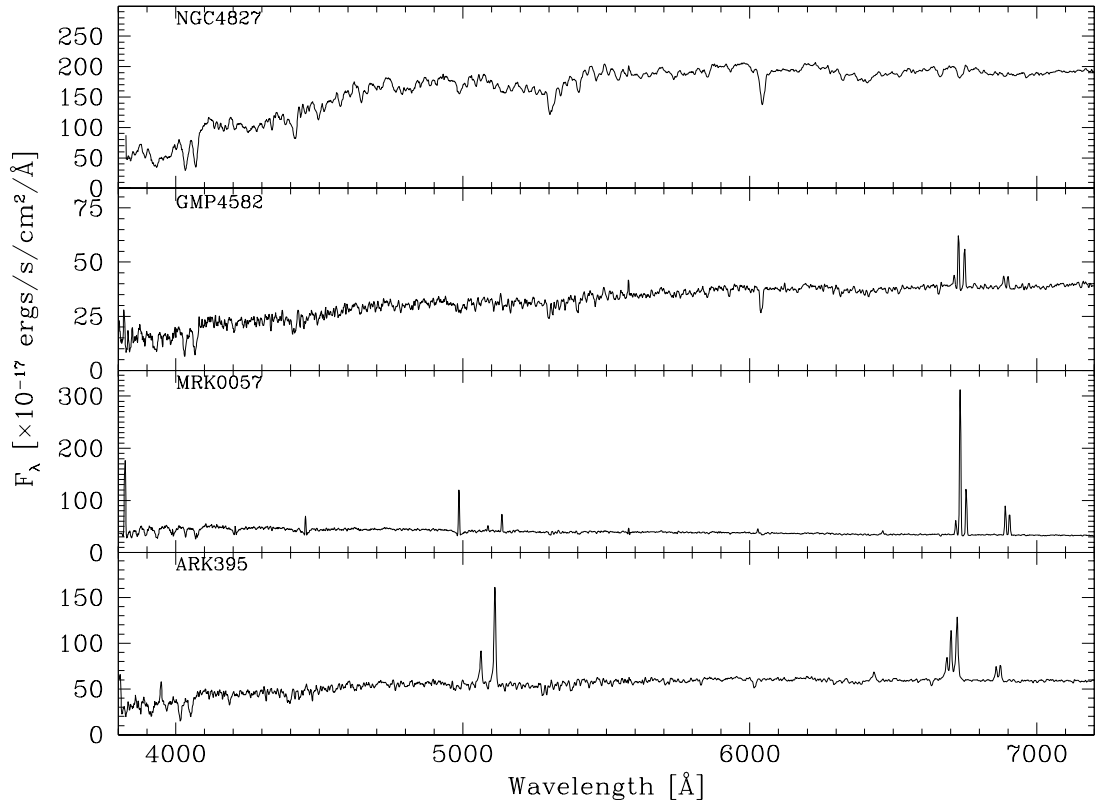}}
\caption{Examples of the four spectral classes in Table \ref{tbl-props}, with galaxy names in top left corner. Top: OSP, or old stellar population; upper middle: SF-cont, or continuous star formation; lower middle: SF-burst, or starburst; bottom: Seyfert. All four spectra are taken from the SDSS and have been boxcar smoothed to the instrumental resolution.\label{fig-specex}}
\end{figure}

\begin{figure}
\figurenum{7}
\epsscale{0.9}
\plotone{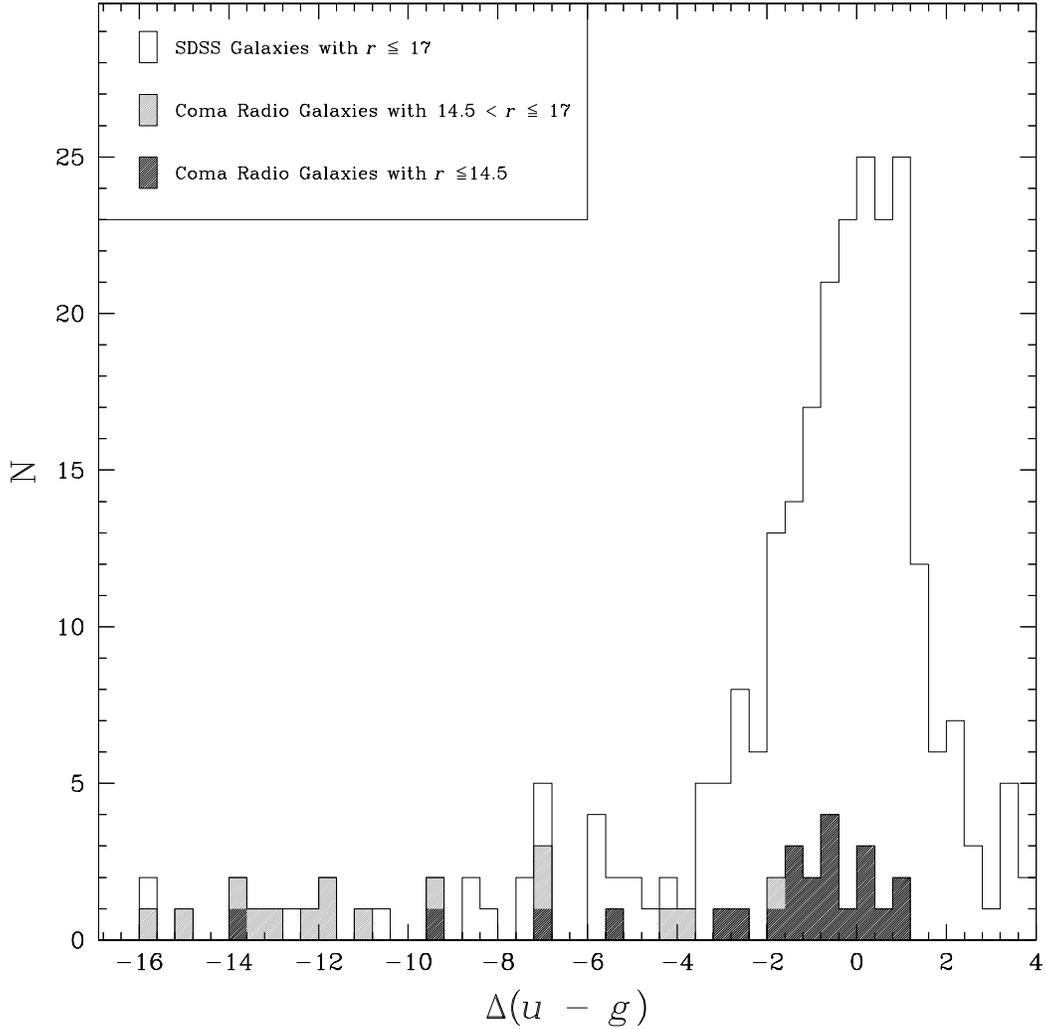}
\caption{Histogram of galaxy colors relative to the derived cluster red sequence. Values are plotted in units of the significance of their deviation from the red sequence, with positive(negative) values being galaxies more red(blue) than would be expected if they laid on the fitted relationship. The open histogram represents all SDSS galaxies with $r \leq 17$ and within the survey area, and has not applied optical spectroscopy to remove foreground and background objects. The shaded histogram represents spectroscopically-confirmed cluster galaxies with $r \leq 17$, with the darker portion for $r \leq 14.5$.\label{fig-cmdhist}}
\end{figure}

\begin{figure}
\figurenum{8}
\epsscale{0.9}
\plotone{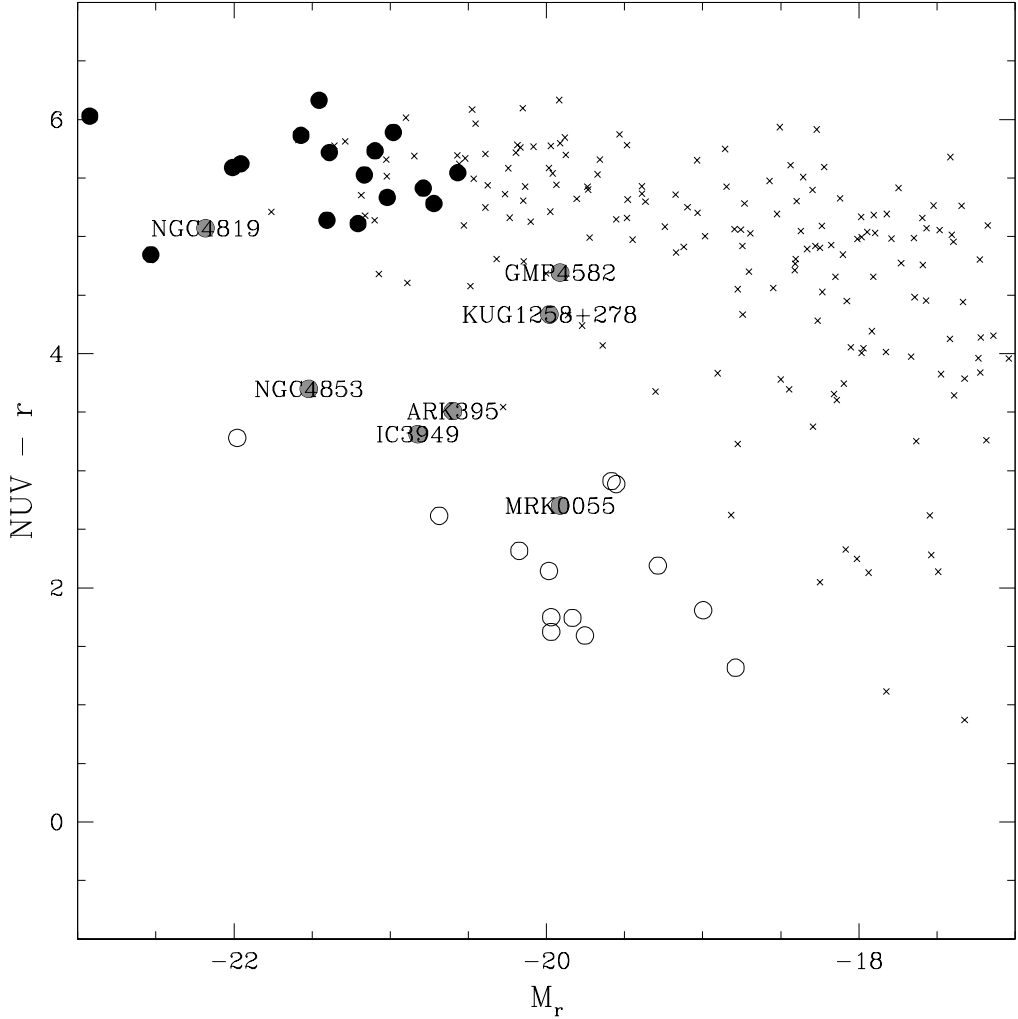}
\caption{$NUV$ and optical color magnitude diagram for galaxies within Coma 1 and Coma 3 regions. Crosses represent GALEX-detected objects within 4\arcsec{} of SDSS galaxies and having a spectroscopic redshift indicating cluster membership. The large circles represent radio-detected members of the Coma cluster, with filled circles being AGN, open circles being star-forming galaxies, and grey circles being the more difficult galaxies to classify (see text; all but NGC4819 and ARK395 are considered to be star-forming). There is a clear separation between radio-detected AGN on the red sequence and radio-detected star-forming galaxies.\label{fig-nuvmr}}
\end{figure}

\begin{figure}
\figurenum{9}
\epsscale{0.9}
\plotone{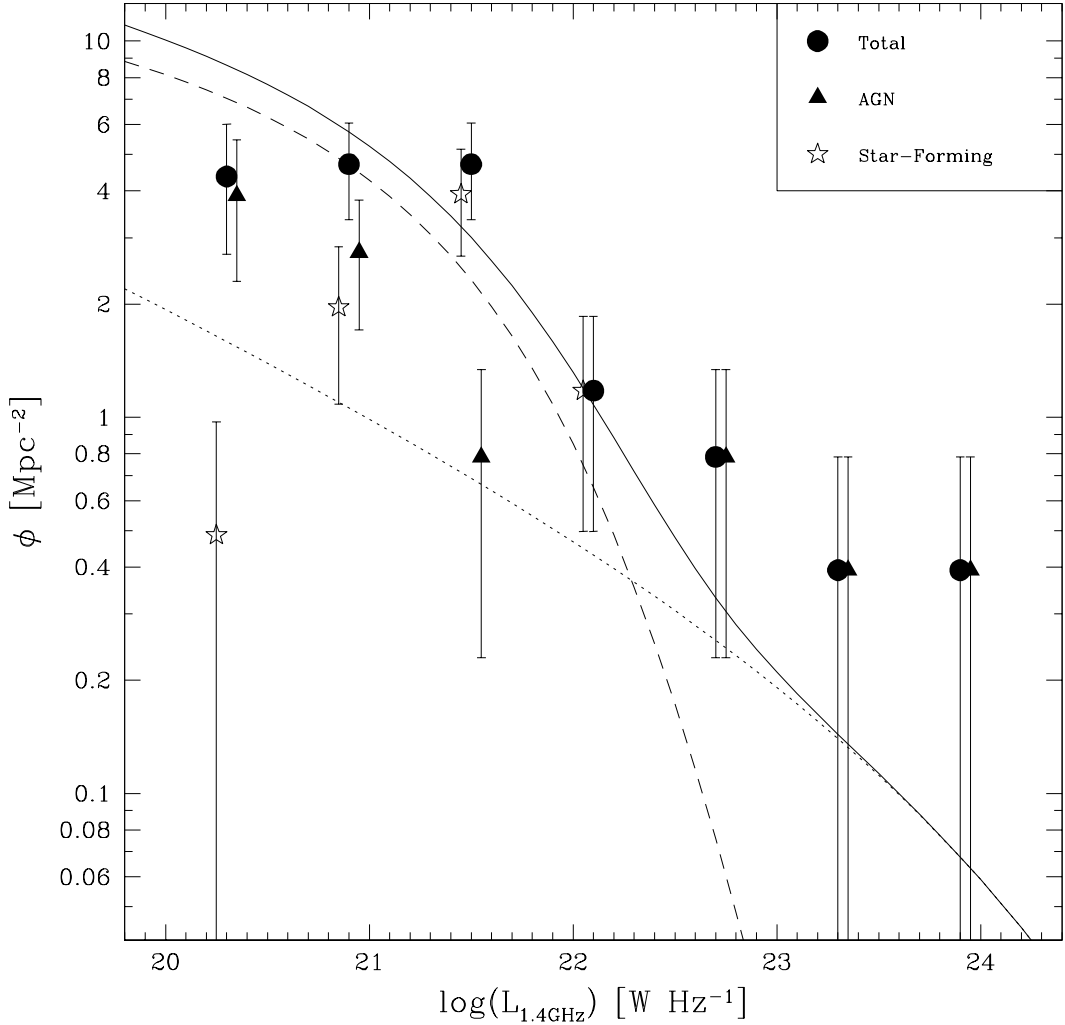}
\caption{Radio luminosity function for the Coma cluster, shown as filled circles. AGN are shown as solid triangles and star-forming galaxies as open stars (each offset slightly from the total RLF for clarity). Approximate completeness limits for unresolved and resolved sources are $\log (L_{1.4GHz})$ of 20.3 and 20.5, respectively (see Figure \ref{fig-ctsbyres}). The dotted line and dashed line represent the functional forms of the RLF from \citet{condon2002} for AGN and star-forming galaxies, respectively, with the relative normalization taken from the cluster RLF work by \citet{lin2007}. The solid line is the total RLF based on these two curves.\label{fig-rlf}}
\end{figure}

\begin{figure}
\figurenum{10}
\epsscale{0.9}
\plotone{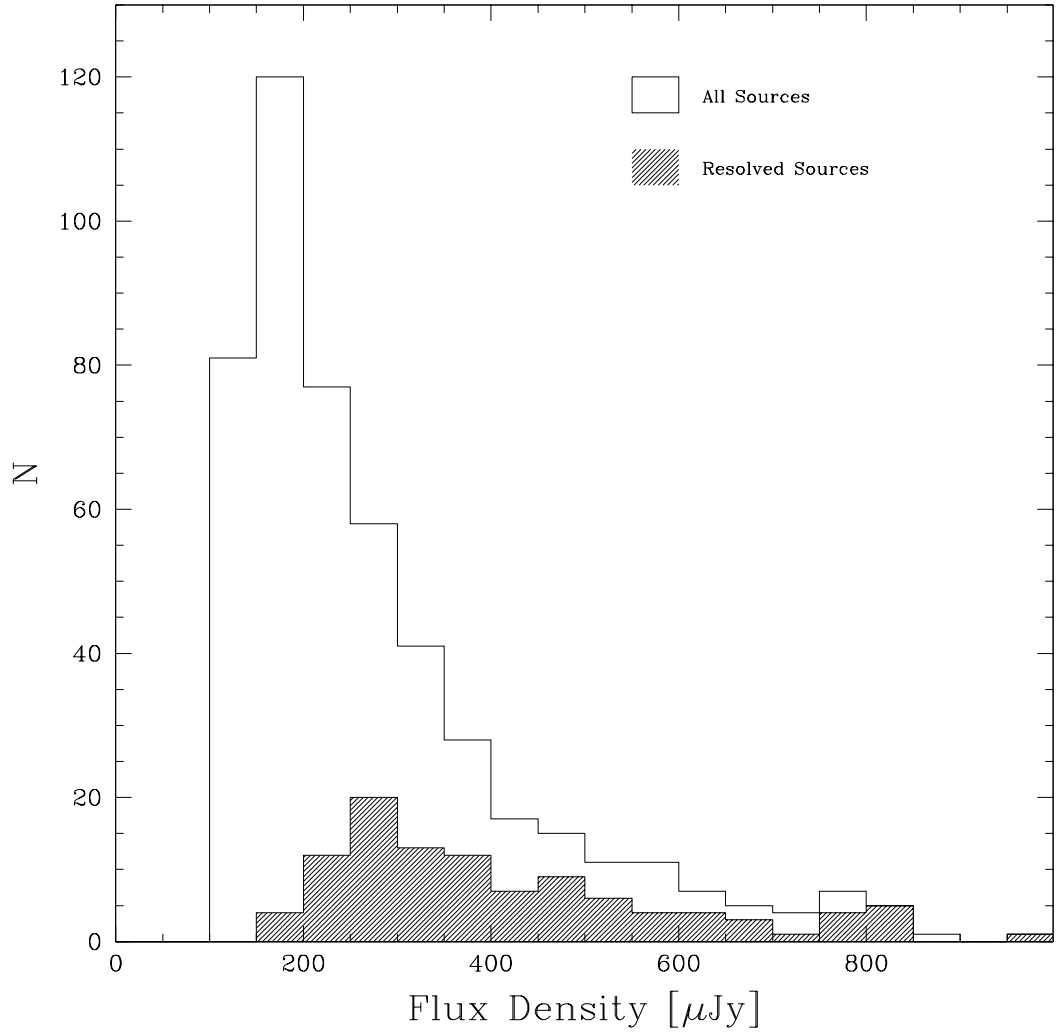}
\caption{Histogram of number of sources by measured flux density. The solid portion represents sources that were resolved.\label{fig-ctsbyres}}
\end{figure}

\begin{figure}
\figurenum{11}
\epsscale{0.9}
\plotone{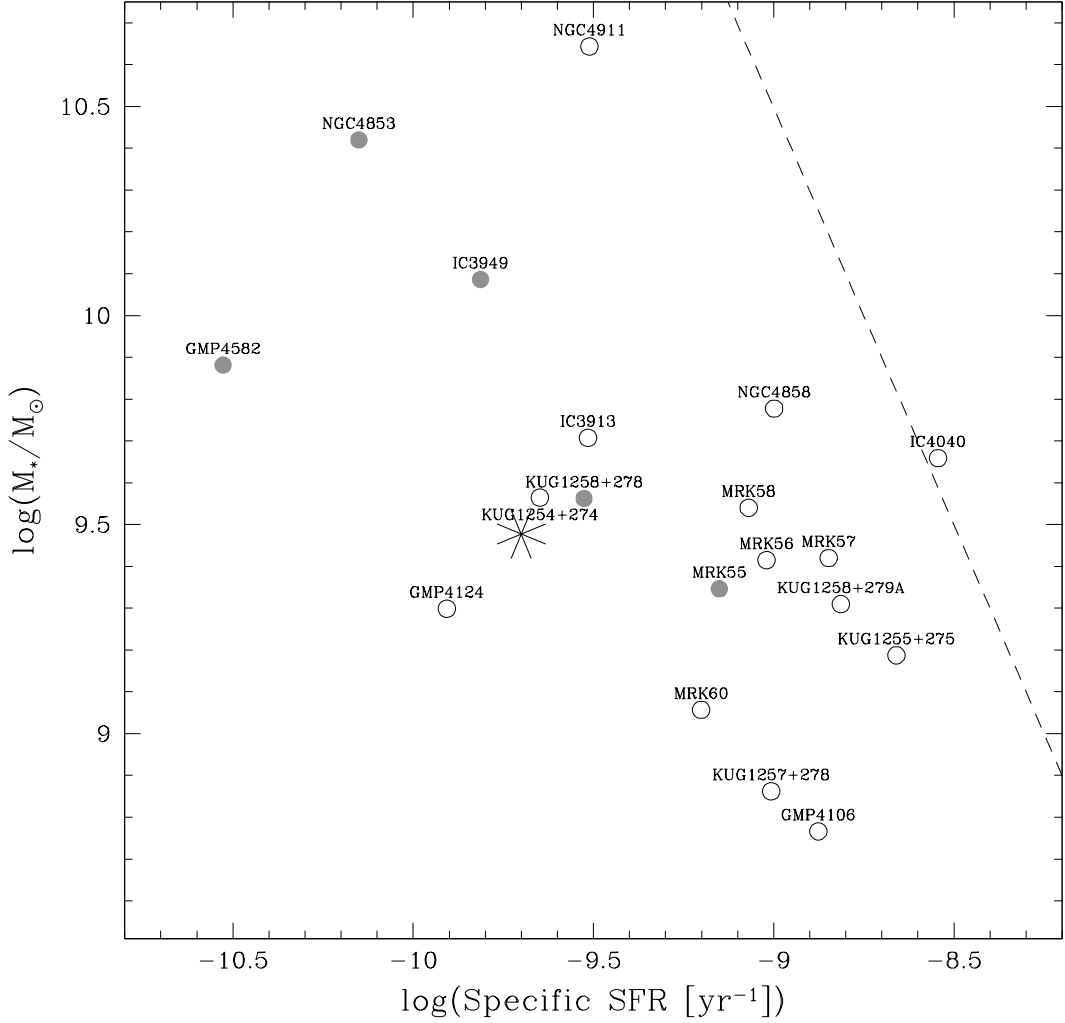}
\caption{Specific SFR for star-forming galaxies. The less certain (possibly transition) objects are noted by grey points. The large asterisk shows the approximate average SSFR for local UV and FIR selected galaxies with stellar mass of $3 \times 10^9$ \myr{} from \citet{buat2007}, and the dashed line is the approximate upper boundary to specific SFR from \citet{feulner2006}. Note the concentration of starburst galaxies with specific SFR around $10^{-9}$ yr$^{-1}$.\label{fig-ssfr}}
\end{figure}

\begin{figure}
\figurenum{12}
\epsscale{0.8}
\plotone{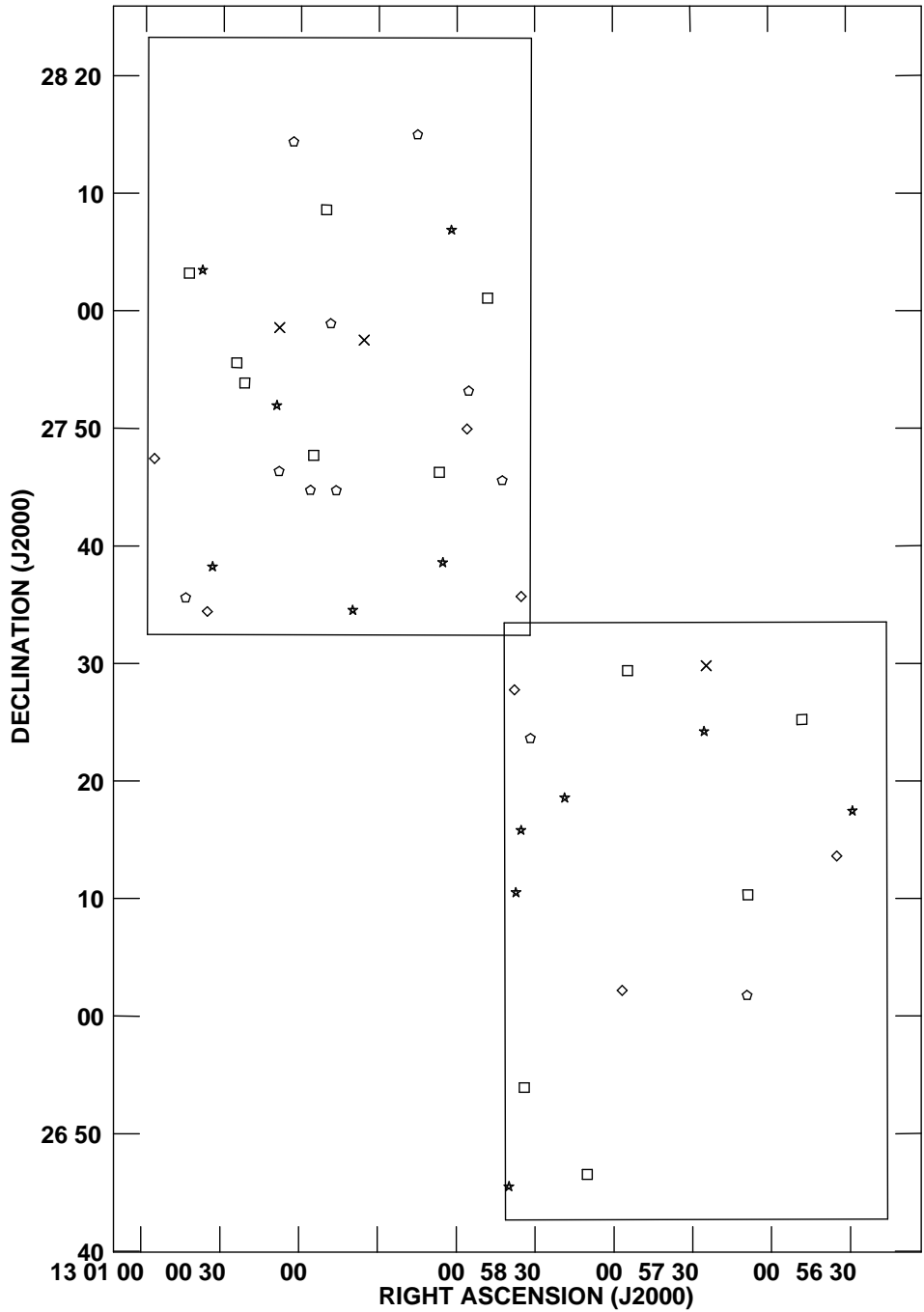}
\caption{Distribution of radio-detected star-forming galaxies and the post-starbursts of P2004. The symbols for the various classes of galaxies are: stars - starburst galaxies; diamonds - non-burst star-forming galaxies; pentagons - blue post-starbursts of P2004; squares - red post-starbursts of P2004; crosses - locations of the three bright Coma ellipticals (NGC4889, NGC4874, and NGC4839). The Coma 1 and Coma 3 boundaries are also depicted.\label{fig-dist}}
\end{figure}

\noindent
{\bf ARK395} The nuclear spectrum of this galaxy is that of a Seyfert 2 (see Figure \ref{fig-specex}). The photometric data are well fit by a synthesized spectrum showing strong star formation, suggesting star formation in the disk of the galaxy along with the active nucleus. Because of the strong active nucleus, we have placed this galaxy in the AGN category but note that star formation likely contributes to the total radio luminosity.

\noindent
{\bf KUG1258+278} This galaxy is slightly bluer than the red sequence fit to the optical data, having $\Delta (u - g) = -4.02$. Similarly, its $(NUV - r) = 4.33$ places it among the transition objects. The NFPS spectrum includes weak H$\beta$ and no noticeable [{\scshape O~iii}], although the synthesized spectrum that best fits the photometry is purely stellar and lacks any emission lines. \citet{miller2001a} indicate that it lies on the FIR/radio correlation, and we place it among the star-forming galaxies.

\noindent
{\bf GMP4582} The photometry for this galaxy is roughly consistent with it lying on the red sequence and thereby being a passive galaxy, with $\Delta (u - g) = -2.00$ and $(NUV - r) = 4.69$. It is also fit better by a de~Vaucouleurs profile. The nuclear spectrum is mainly that of an older stellar population, although weak emission from [{\scshape O~iii}], H$\alpha$, and [{\scshape N~ii}] are detected. The resulting line ratios, once corrected for underlying stellar Balmer absorption, place this galaxy just barely within the star-forming class. The radio luminosity translates to a SFR of 0.2 \myr, consistent with the interpretation that this is a moderately massive galaxy dominated by older stellar populations but with slight star formation activity.

\noindent
{\bf MRK0055} NED indicates this galaxy is a starburst, which is consistent with its blue colors in both the optical and UV. However, we note that its nuclear spectrum was classified as a LINER in \citet{miller2002}. It has been placed among the star-forming galaxies in our analysis.

\begin{figure}
\figurenum{A.1}
\epsscale{0.7}
\rotatebox{270}{\plotone{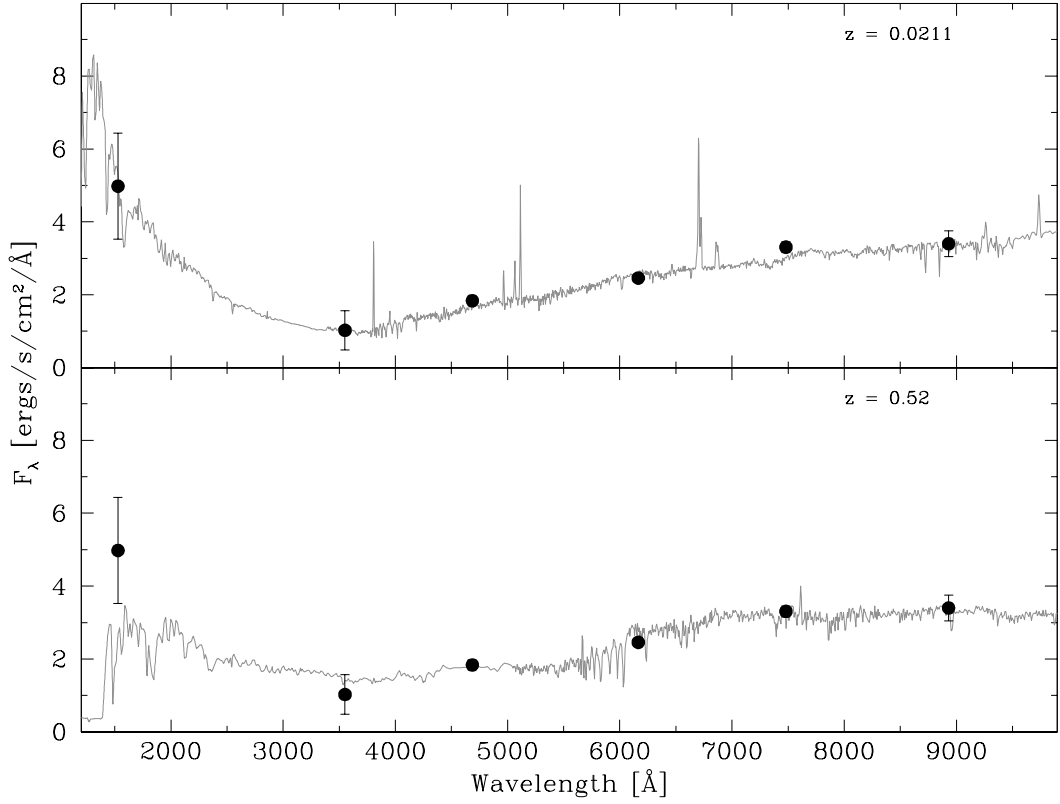}}
\caption{K\_CORRECT best fitting synthesized spectra for J12591536+2746052, under the assumptions that it resides in Coma (top) and that it lies at $z=0.52$ (bottom). The photometric data points and their errors (1$\sigma$) are plotted in black.\label{fig-nonclus}}
\end{figure}

\noindent
{\bf J12591536+2746052} As noted in Section \ref{sec-optspec}, the velocity measurement for this galaxy was taken from \citet{biviano1995}, who indicated that the reliability of this velocity was low because it was based on only two spectral features. It is best fit by an exponential profile, yet would lie on the Coma red sequence ( $\Delta (u - g) = -0.43$). The GALEX $NUV$ magnitude is quite bright relative to the other bands, and indeed is flagged as a probable extraction error. Our attempts to fit the photometric data underscored this, as inclusion of the $NUV$ produced synthetic spectral fits that failed to match most of the photometry. Ignoring the GALEX data and assuming the cluster redshift was correct did produce a reasonable fit to the photometry, that of an unusually reddened star-forming galaxy (see Figure \ref{fig-nonclus}). Table \ref{tbl-props} includes the stellar mass associated with this fit. As a second interpretation, we note that K\_CORRECT produced a photometric redshift greater than one for this galaxy and its photometric redshifts from the SDSS DR6 were $0.59\pm0.09$ and $0.52\pm0.09$. We consequently also used K\_CORRECT to fit the photometric data under the assumption that $z=0.52$ and found that a post-starburst spectrum at this redshift is also a good match. The galaxy's stellar mass associated with this fit was $9 \times 10^{10}$ M$_\odot$. Assuming a synchrotron spectral index, the rest frame radio luminosity of this galaxy should it lie at $z=0.52$ is $1.6 \times 10^{23}$ \whz. While this may seem quite large, we note that \citet{smail1999} detected several post-starburst galaxies with comparable radio luminosities in their study of the Cl0939+4713 cluster at $z=0.41$. Furthermore, if the $z=0.0211$ fit were correct, the multiple emission lines should have led to a higher reported spectrum quality in \citet{biviano1995}. We consequently assumed this galaxy was a background object and not a cluster member. Where pertinent, we have indicated how this assumption affected our interpretations. 

% Castander et al. include NGC4842A (fiber 219), NGC4827 (248), MRK58 (168), IC3976 (142), KUG1258+278 (106),
% MRK56 (126), MRK55 (209), KUG1258+279A (98), KUG1255+275 (127), and GMP4106 (115). Indicate that:
%  NGC4827-IC3976 are pure absorption spectra
%  MRK58-MRK56-KUG1255+275-GMP4106 are strong emission line spectra
%  MRK55-KUG1258+279A emission lines plus absorption lines
%  NGC4842A-KUG1258+278 are absorption lines plus emission lines

% Note SFR for probable background is 0.12 from radio if in Coma, 0.0065 from FUV (did not account for 
% extinction within the galaxy itself).
% Photo-z are described in DR5 paper, section 4

\clearpage

\clearpage

\begin{deluxetable}{r r r r r r r l}
\tablecolumns{8}
\tablecaption{Cluster Member Radio-Detected Galaxies\label{tbl-mems}}
\tablewidth{0pt}
\tablehead{
\colhead{RA} & \colhead{Dec} & \colhead{$r$ mag} &
\colhead{$S_{1.4GHz}$} & \colhead{$\log(L_{1.4GHz})$} & \colhead{$cz$} & \colhead{Ref} 
& \colhead{Name} \\
\colhead{(J2000)} & \colhead{(J2000)} & \colhead{} & \colhead{[$\mu$Jy]} 
& \colhead{[\whz]} & \colhead{[\kms]} & \colhead{} & \colhead{} 
}
\startdata
13 00 08.14 & +27 58 37.2 & 11.974 &   1194$\pm$\phn43 & 21.1520 & 6495 &  9 &      NGC4889 \\
12 59 35.71 & +27 57 33.5 & 12.102 & 206290$\pm$174    & 23.3893 & 7174 &  1 &      NGC4874 \\
12 57 24.36 & +27 29 52.1 & 12.495 &  77027$\pm$478    & 22.9616 & 7346 &  1 &      NGC4839 \\
12 56 27.86 & +26 59 14.6 & 12.849 &    304$\pm$\phn69 & 20.5579 & 6463 &  6 &      NGC4819 \\
12 56 43.53 & +27 10 43.7 & 13.015 &  63740$\pm$603    & 22.8794 & 7593 &  1 &      NGC4827 \\
13 00 56.06 & +27 47 27.2 & 13.041 &  19085$\pm$259    & 22.3556 & 7939 &  6 &      NGC4911 \\
13 00 17.93 & +28 12 08.6 & 13.077 &    492$\pm$\phn59 & 20.7670 & 8492 & 12 &      NGC4895 \\
12 59 19.87 & +28 05 03.5 & 13.457 &   2685$\pm$\phn42 & 21.5040 & 4700 & 11 &      NGC4865 \\
12 58 35.19 & +27 35 47.0 & 13.508 &   2614$\pm$\phn98 & 21.4923 & 7712 &  1 &      NGC4853 \\
12 59 13.13 & +27 58 37.2 & 13.575 &   1007$\pm$\phn43 & 21.0781 & 6841 & 12 &      NGC4864 \\
12 59 23.37 & +27 54 41.8 & 13.626 & 402020$\pm$697    & 23.6793 & 6859 & 11 &      NGC4869 \\
13 00 51.55 & +28 02 34.4 & 13.640 &    421$\pm$\phn29 & 20.6993 & 8793 & 12 &       IC4051 \\
13 00 17.69 & +27 57 19.1 & 13.822 &    368$\pm$\phn23 & 20.6409 & 6661 &  5 &      NGC4898 \\
13 00 48.65 & +28 05 26.5 & 13.863 &    293$\pm$\phn29 & 20.5419 & 6940 & 12 &       IC4045 \\
12 57 35.86 & +27 29 35.5 & 13.929 &   1297$\pm$\phn58 & 21.1880 & 7343 &  3 &     NGC4842A \\
13 00 42.77 & +27 58 16.7 & 14.009 &    272$\pm$\phn27 & 20.5096 & 6392 & 11 &       IC4042 \\
12 59 01.82 & +28 13 31.1 & 14.050 &    438$\pm$\phn28 & 20.7165 & 8028 &  1 &      GMP3818 \\
12 58 55.97 & +27 50 00.2 & 14.215 &   2637$\pm$\phn85 & 21.4961 & 7481 &  7 &       IC3949 \\
12 59 15.24 & +27 58 14.5 & 14.242 &    131$\pm$\phn23 & 20.1923 & 4859 & 10 &      NGC4867 \\
12 59 34.13 & +27 56 48.5 & 14.305 &    210$\pm$\phn54 & 20.3972 & 7183 &  7 &      NGC4872 \\
12 59 05.30 & +27 38 39.8 & 14.344 &   4158$\pm$117    & 21.6939 & 5554 &  5 &      MRK0058 \\
12 56 51.17 & +26 53 56.0 & 14.428 &   1856$\pm$\phn80 & 21.3436 & 6210 &  1 &       ARK395 \\
12 59 29.40 & +27 51 00.4 & 14.462 &    168$\pm$\phn23 & 20.3003 & 6865 & 11 &       IC3976 \\
12 56 28.56 & +27 17 28.7 & 14.854 &   2191$\pm$173    & 21.4157 & 7514 &  1 &       IC3913 \\
13 00 35.69 & +27 34 27.1 & 15.048 &   1529$\pm$102    & 21.2594 & 5112 & 11 &  KUG1258+278 \\
12 58 35.33 & +27 15 52.9 & 15.057 &   3494$\pm$104    & 21.6183 & 7368 &  5 &      MRK0056 \\
13 00 37.87 & +28 03 29.2 & 15.061 &  18309$\pm$160    & 22.3375 & 7850 &  5 &       IC4040 \\
12 58 37.27 & +27 10 35.8 & 15.068 &   5254$\pm$136    & 21.7955 & 7665 &  1 &      MRK0057 \\
12 57 56.66 & +27 02 15.0 & 15.115 &    318$\pm$\phn63 & 20.5775 & 7428 &  1 &      GMP4582 \\
12 57 25.25 & +27 24 16.6 & 15.119 &   2206$\pm$\phn68 & 21.4186 & 4857 &  8 &      MRK0055 \\
12 59 02.06 & +28 06 56.6 & 15.200 &   8452$\pm$137    & 22.0020 & 9422 &  1 &      NGC4858 \\
13 00 33.67 & +27 38 16.1 & 15.275 &   4404$\pm$136    & 21.7189 & 7476 &  5 & KUG1258+279A \\
12 56 34.63 & +27 13 39.4 & 15.442 &   1160$\pm$485\tablenotemark{a} & 21.1395 & 7197 &  4 & KUG1254+274  \\
13 00 09.14 & +27 51 59.4 & 15.473 &   1009$\pm$\phn42 & 21.0789 & 5293 &  7 &      MRK0060 \\
12 59 39.82 & +27 34 35.4 & 15.740 &   1007$\pm$140    & 21.0781 & 5006 &  1 &  KUG1257+278 \\
12 58 37.82 & +27 27 50.4 & 15.924 &    347$\pm$\phn78 & 20.6154 & 6303 &  1 &      GMP4124 \\
12 58 18.63 & +27 18 38.9 & 16.042 &   4739$\pm$113    & 21.7507 & 7447 &  4 &  KUG1255+275 \\
12 58 39.93 & +26 45 34.2 & 16.238 &   1092$\pm$172    & 21.1132 & 7467 &  4 &      GMP4106 \\
12 59 15.36 & +27 46 05.2 & 20.194 &    177$\pm$\phn25 & 20.3230 & 6315 &  2 &      \nodata \\
\enddata

\tablenotetext{a}{Faint and extended radio source added after detection in convolved radio images (see Section \ref{sec-adds}).}

\tablecomments{Galaxies are ordered by their extinction-corrected $r$ magnitude, with
positions signifying the coordinates of the SDSS optical counterpart to each radio source.
References for the velocity measurements are coded as follows: (1) \citet{adelman2008}, 
(2) \citet{biviano1995}, (3) \citet{caldwell1993}, (4) \citet{castander2001}, 
(5) \citet{devaucoleurs1991}, (6) \citet{falco1999}, (7) Marzke et al., in preparation, 
(8) \citet{miller2001a}, (9) \citet{moore2002}, (10) \citet{mueller1999}, 
(11) \citet{smith2004}, (12) \citet{smith2000}. Source names include NGC for New General 
Catalog, IC for Index Catalog, MRK for Markaryan objects, KUG for Kiso Ultraviolet Galaxy 
catalog, and GMP for \citet{godwin1983}.}

\end{deluxetable}

\begin{deluxetable}{r r r r r c r l r l}
\tablecolumns{10}
\tabletypesize{\scriptsize}
\tablecaption{Additional Properties of Coma Radio-Detected Galaxies\label{tbl-props}}
\tablewidth{0pt}
\tablehead{
\colhead{RA(J2000)} & \colhead{Dec(J2000)} & \colhead{Profile} & \colhead{$\Delta (u-g)$} & 
\colhead{$NUV - r$} & \colhead{SpecClass} & \colhead{$\log (M_*/M_\odot)$} & \colhead{Type} & 
\colhead{SFR} & \colhead{Name}
}
\startdata
13 00 08.14 & +27 58 37.2 & D &  -1.49 &    7.39\tablenotemark{a} & OSP\tablenotemark{b} & 11.06 & AGN & \nodata & NGC4889 \\
12 59 35.71 & +27 57 33.5 & D &  -1.90 &    6.03 &      OSP & 11.07                 & AGN & \nodata & NGC4874 \\
12 57 24.36 & +27 29 52.1 & D &   0.67 &    4.85 &      OSP & 10.94                 & AGN & \nodata & NGC4839 \\
12 56 27.86 & +26 59 14.6 & D &  -2.45 &    5.07 &  \nodata & 10.64                 & AGN & \nodata & NGC4819 \\
12 56 43.53 & +27 10 43.7 & D &   0.06 &    5.59 &      OSP & 10.74                 & AGN & \nodata & NGC4827 \\
13 00 56.06 & +27 47 27.2 & D &  -3.20 &    3.28\tablenotemark{c} & \nodata & 10.64 &  SF &    13.4 & NGC4911 \\
13 00 17.93 & +28 12 08.6 & D &  -1.51 &    5.62 & OSP\tablenotemark{b} & 10.82     & AGN & \nodata & NGC4895 \\
12 59 19.87 & +28 05 03.5 & D &  -0.43 &    5.86 & OSP\tablenotemark{b} & 10.16     & AGN & \nodata & NGC4865 \\
12 58 35.19 & +27 35 47.0 & D &  -9.35 &    3.70 &  SF-cont & 10.42                 &  SF &     1.8 & NGC4853 \\
12 59 13.13 & +27 58 37.2 & D &  -0.89 &    6.17 & OSP\tablenotemark{b} & 10.43     & AGN & \nodata & NGC4864 \\
12 59 23.37 & +27 54 41.8 & D &   0.29 &    5.14 & OSP\tablenotemark{b} & 10.39     & AGN & \nodata & NGC4869 \\
13 00 51.55 & +28 02 34.4 & D &  -0.64 &    5.27 & OSP\tablenotemark{b} & 10.63     & AGN & \nodata &  IC4051 \\
13 00 17.69 & +27 57 19.1 & D &  -1.56 &    5.11 & OSP\tablenotemark{b} & 10.28     & AGN & \nodata & NGC4898 \\
13 00 48.65 & +28 05 26.5 & D &  -0.80 &    5.53 & OSP\tablenotemark{b} & 10.33     & AGN & \nodata &  IC4045 \\
12 57 35.86 & +27 29 35.5 & D &   1.09 &    5.73 &  \nodata & 10.36                 & AGN & \nodata & NGC4842A \\
13 00 42.77 & +27 58 16.7 & D &  -0.37 &    5.33 & OSP\tablenotemark{b} & 10.17     & AGN & \nodata &  IC4042 \\
12 59 01.82 & +28 13 31.1 & D &  -0.94 &    5.89 &      OSP & 10.40                 & AGN & \nodata & GMP3818 \\
12 58 55.97 & +27 50 00.2 & E &  -5.27 &    3.31 &  SF-cont & 10.09                 &  SF &     1.8 &  IC3949 \\
12 59 15.24 & +27 58 14.5 & D &   0.22 &    5.41 &  \nodata &  9.85                 & AGN & \nodata & NGC4867 \\
12 59 34.13 & +27 56 48.5 & D &  -0.43 &    5.28 &      OSP & 10.16                 & AGN & \nodata & NGC4872 \\
12 59 05.30 & +27 38 39.8 & D & -13.70 &    2.62 & \tablenotemark{d} &  9.54        &  SF &     2.9 & MRK0058 \\
12 56 51.17 & +26 53 56.0 & E &  -7.04 &    3.51 &  Seyfert &  9.86                 & AGN & \nodata &  ARK395 \\
12 59 29.40 & +27 51 00.4 & D &   0.95 &    5.55 & OSP\tablenotemark{b} & 10.09     & AGN & \nodata &  IC3976 \\
12 56 28.56 & +27 17 28.7 & E &  -9.27 &    2.32 & SF-burst &  9.71                 &  SF &     1.5 &  IC3913 \\
13 00 35.69 & +27 34 27.1 & E &  -4.02 &    4.33 & SF-cont\tablenotemark{b} &  9.56 &  SF &     1.1 & KUG1258+278 \\
12 58 35.33 & +27 15 52.9 & D & -15.90 &    2.14 & \tablenotemark{d} &  9.42        &  SF &     2.5 & MRK0056 \\
13 00 37.87 & +28 03 29.2 & E & -11.70 &    1.75 & \tablenotemark{d} &  9.66        &  SF &    12.8 &  IC4040 \\
12 58 37.27 & +27 10 35.8 & E & -12.90 &    1.62 & SF-burst\tablenotemark{d} & 9.42 &  SF &     3.7 & MRK0057 \\
12 57 56.66 & +27 02 15.0 & D &  -2.00 &    4.69 &  SF-cont &  9.88                 &  SF &     0.2 & GMP4582 \\
12 57 25.25 & +27 24 16.6 & D & -10.90 &    2.70 & \tablenotemark{d} &  9.35        &  SF &     1.5 & MRK0055 \\
12 59 02.06 & +28 06 56.6 & E & -13.80 &    1.74 & SF-burst\tablenotemark{d} & 9.79 &  SF &     5.9 & NGC4858 \\
13 00 33.67 & +27 38 16.1 & E & -15.10 &    1.59 & SF-burst\tablenotemark{b,d} & 9.31 &  SF &     3.1 & KUG1258+279A \\
12 56 34.63 & +27 13 39.4 & E &  -7.10 &    2.91 & \nodata  &  9.56                 &  SF &     0.8 & KUG1254+274 \\
13 00 09.14 & +27 51 59.4 & E &  -7.16 &    2.89 & SF-burst\tablenotemark{d} & 9.06 &  SF &     0.7 & MRK0060 \\
12 59 39.82 & +27 34 35.4 & E & -12.00 &    2.19 & SF-burst &  8.86                 &  SF &     0.7 & KUG1257+278 \\
12 58 37.82 & +27 27 50.4 & E &  -3.91 & \tablenotemark{e} & SF-cont &  9.30        &  SF &     0.2 & GMP4124 \\
12 58 18.63 & +27 18 38.9 & E & -11.70 &    1.81 & \tablenotemark{d} &  9.19        &  SF &     3.3 & KUG1255+275 \\
12 58 39.93 & +26 45 34.2 & E & -13.20 &    1.32 & \tablenotemark{d} &  8.77        &  SF &     0.8 & GMP4106 \\
12 59 15.36 & +27 46 05.2 & E &  -0.43 & \tablenotemark{f} & \nodata &  7.79     & \nodata & \nodata & \nodata \\
\enddata

\tablenotetext{a}{Presumably the very red $NUV - r$ color is due to errors in photometry caused by the extended nature of this source.}

\tablenotetext{b}{Based on NFPS spectrum.}

\tablenotetext{c}{The GALEX $NUV$ magnitude for this galaxy was approximated as the sum of a pair of sources, as
the automated photometry of the SDSS divided the galaxy into two components.}

\tablenotetext{d}{NED indicates a starburst or {\scshape H~ii} classification for this galaxy.}

\tablenotetext{e}{There is a GALEX FUV detection for this galaxy, but not an NUV detection.}

\tablenotetext{f}{There is a GALEX NUV detection for this galaxy, but it is flagged. The $(NUV - r)$ implied
by this measurement is 0.32, and this anomalously high $NUV$ flux can not be matched with normal galaxy 
templates (see text).}

\tablecomments{``Profile'' refers to the better fitting profile as determined by the SDSS, with ``D'' for de~Vaucoleurs and ``E'' for exponential, generally representing elliptical and spiral galaxies, respectively. Abbreviations for spectral classes are: OSP for old stellar population (absorption line spectra lacking significant emission lines), SF-cont for spectra indicative of low levels of star formation, SF-burst for spectra indicative of high levels of star formation, and Seyfert for emission line AGN. Masses are determined from the K\_CORRECT \citep{blanton2007} fits to the SDSS and GALEX photometric data, and represent the total stellar masses of the galaxies. The ``Type'' column refers to the assumed activity class of the galaxy on the basis of all available data, and star formation rates (SFR) are in units of \myr{} for star-forming galaxies only. }

\end{deluxetable}

\begin{deluxetable}{r r r r r}
\tablecolumns{5}
\tablecaption{Mean Pair Separations\label{tbl-pairs}}
\tablewidth{0pt}
\tablehead{
\colhead{} & \colhead{Radio SB} & \colhead{Radio SF} & \colhead{Blue PSB} & \colhead{Red PSB} \\
\colhead{} & \colhead{(N=12)}     & \colhead{(N=7)}      & \colhead{(N=11)}     & \colhead{(N=12)}
}
\startdata
Radio SB & 721\arcsec & 708\arcsec & 670\arcsec & 570\arcsec \\
Radio SF & 511\arcsec & 870\arcsec & 472\arcsec & 610\arcsec \\
Blue PSB & 558\arcsec & 713\arcsec & 665\arcsec & 450\arcsec \\
Red PSB  & 415\arcsec & 727\arcsec & 675\arcsec & 622\arcsec \\
\enddata

\tablecomments{The four classes correspond to: Radio SB = radio-detected starburst galaxy, Radio SF = radio-detected galaxy with more gradual star formation histories, Blue PSB = blue post-starburst of P2004, Red PSB = red post-starburst of P2004. The number of objects in each class is indicated in the second row of the column headers. Values are read along rows, for example the mean separation from a radio-detected starburst to its nearest blue post-starburst is 670\arcsec.}
\end{deluxetable}

\begin{deluxetable}{r r r r r}
\tablecolumns{5}
\tablecaption{Monte Carlo Pair Separation Results\label{tbl-mcpairs}}
\tablewidth{0pt}
\tablehead{
\colhead{} & \colhead{Radio SB} & \colhead{Radio SF} & \colhead{Blue PSB} & \colhead{Red PSB} 
}
\startdata
\cutinhead{Holding cluster-centric radius fixed:}
Radio SB & 84.0 & 17.8 & 59.3 & 59.0 \\
Radio SF & 14.3 & 45.6 & 12.3 & 48.7 \\
Blue PSB & 60.4 & 56.2 & 20.2 & 40.0 \\
Red PSB  & 14.1 & 49.2 & 91.4 & 77.0 \\
\cutinhead{Fully randomized:}
Radio SB & 86.4 & 52.3 & 71.6 & 62.2 \\
Radio SF & 37.8 & 55.6 & 10.6 & 64.0 \\
Blue PSB & 64.1 & 55.7 & 63.8 & 26.8 \\
Red PSB  &  9.4 & 57.7 & 77.9 & 61.0 \\
\enddata

\tablecomments{The four classes and reading of the data are the same as in Table \ref{tbl-pairs}. Values presented are the percentage of MC simulations which produced smaller mean separations than the actual data, hence lower numbers indicate more significant clustering is present in the observed Coma populations.}
\end{deluxetable}

\end{document}